\documentclass[twocolumn,aps,superscriptaddress,multicol,amsmath,amssymb,longbibliography]{revtex4-2}
\usepackage[utf8]{inputenc}
\usepackage[normalem]{ulem}
\usepackage{amsmath}
\usepackage{siunitx}
\usepackage{graphicx}
\usepackage{sidecap}
\usepackage[dvipsnames]{xcolor}
\graphicspath{{.}{./figures-main/}}
\usepackage{nameref,zref-xr}
\zxrsetup{toltxlabel}
\usepackage[colorlinks,linkcolor=blue,citecolor=blue,anchorcolor=blue]{hyperref}
\usepackage{cleveref}
\begin{document}

\title{Stacking of charge-density waves in 2H-NbSe$_2$ bilayers}

\author{F. Cossu}
\email[F. Cossu: ]{fabrizio.cossu@york.ac.uk}
\affiliation{Department of Physics and Institute of Quantum Convergence and Technology, Kangwon National University -- Chuncheon, 24341, Korea}
\affiliation{Asia Pacific Center for Theoretical Physics -- Pohang, Gyeongbuk, 37673, Korea}
\affiliation{Department of Physics, School of Natural and Computing Sciences, University of Aberdeen, Aberdeen, AB24 3UE, UK}
\affiliation{School of Physics, Engineering and Technology, University of York, Heslington, York YO10 5DD, United Kingdom}

\author{D. Nafday}
\affiliation{Asia Pacific Center for Theoretical Physics -- Pohang, Gyeongbuk, 37673, Korea}
\affiliation{Department of Applied Physics, School of Engineering Sciences, KTH Royal Institute of Technology, AlbaNova University Center, Stockholm, SE-10691, Sweden}

\author{K. Palot\'{a}s}
\affiliation{Institute for Solid State Physics and Optics, HUN-REN Wigner Research Center for Physics, H-1525 Budapest, Hungary}
\affiliation{HUN-REN-SZTE Reaction Kinetics and Surface Chemistry Research Group, University of Szeged, H-6720 Szeged, Hungary}

\author{M. Biderang}
\affiliation{Department of Physics, University of Toronto, Toronto, Ontario, M5S 1A7, Canada}

\author{H.-S. Kim}
\affiliation{Department of Physics and Institute of Quantum Convergence and Technology, Kangwon National University -- Chuncheon, 24341, Korea}

\author{A. Akbari}
\affiliation{Asia Pacific Center for Theoretical Physics -- Pohang, Gyeongbuk, 37673, Korea}
\affiliation{Bejing Institute of Mathematical Sciences and Applications (BIMSA), Huairou District, Beijing, 101408, China}
\affiliation{Institut für Theoretische Physik III, Ruhr-Universität Bochum, D-44801 Bochum, Germany}

\author{I. {Di Marco}}
\email[I. {Di Marco}: ]{igor.dimarco@umk.pl}
\affiliation{Asia Pacific Center for Theoretical Physics -- Pohang, Gyeongbuk, 37673, Korea}
\affiliation{Department of Physics and Astronomy, Uppsala University, Box 516, SE-75120, Uppsala, Sweden}
\affiliation{Institute of Physics, Nicolaus Copernicus University, 87-100 Toru\'n, Poland}

\date{\today}

\begin{abstract}
    We employ \textit{ab-initio} electronic structure calculations to investigate the charge-density waves and periodic lattice distortions in bilayer 2H-NbSe$_2$. 
    We demonstrate that the vertical stacking can give rise to a variety of patterns that may lower the symmetry of the charge-density waves exhibited separately by the two composing 1H-NbSe$_2$ monolayers.
    The general tendency to a spontaneous symmetry breaking observed in the ground state and the first excited states is shown to originate from a non-negligible inter-layer coupling.
    Simulated images for scanning tunnelling microscopy as well as geometric structure factors show signatures of the different stacking orders. This may not only be useful to reinterpret past experiments on surfaces and thin films, but may also be exploited to devise ad-hoc experiments for the investigation of the stacking order in 2H-NbSe$_2$.
    We anticipate that our analysis does not only apply to the 2H-NbSe$_2$, but is also relevant for thin films and bulk, whose smallest centro-symmetric component is indeed the bilayer. 
    Finally, our results illustrate clearly that the vertical stacking is not only important for 1T structures, as exemplified by the metal-to-insulator transition observed in 1T-TaS$_2$, but seems to be a general feature of metallic layered transition metal dichalcogenides as well. 
\end{abstract}


\maketitle 

\section{Introduction}
Transition metal dichalcogenides (TMD) are a class of layered materials that have been under intense scrutiny for several decades~\cite{chowdhury_t-ChemRev2020}. The weak inter-layer coupling, originating from the van der Waals (vdW) interaction, makes them relatively easy to exfoliate~\cite{joensen-MResBull1986}, allowing for the systematic study of thickness-dependent quantum phenomena. 
This is very important in the study of charge-density waves (CDWs)~\cite{Rossnagel_2011,xu_zq-NanoT2021}, as understanding their evolution from monolayer to bulk may clarify the precise mechanisms driving their formation, which are still heavily debated~\cite{zhu-PNAS2015,PhysRevResearch.5.013218,lin_dj-ncomm2020}.
In turn, this may help to understand the connection between CDWs, their dimensionality and other phenomena driven by strong electronic correlations, as e.g. superconductivity and magnetism~\cite{das_s-npjCM2023,lin_dj-ncomm2020}.

\begin{figure*}
\centering
\includegraphics[trim= 0cm 1.5cm 0cm 0cm, width=0.24\linewidth]{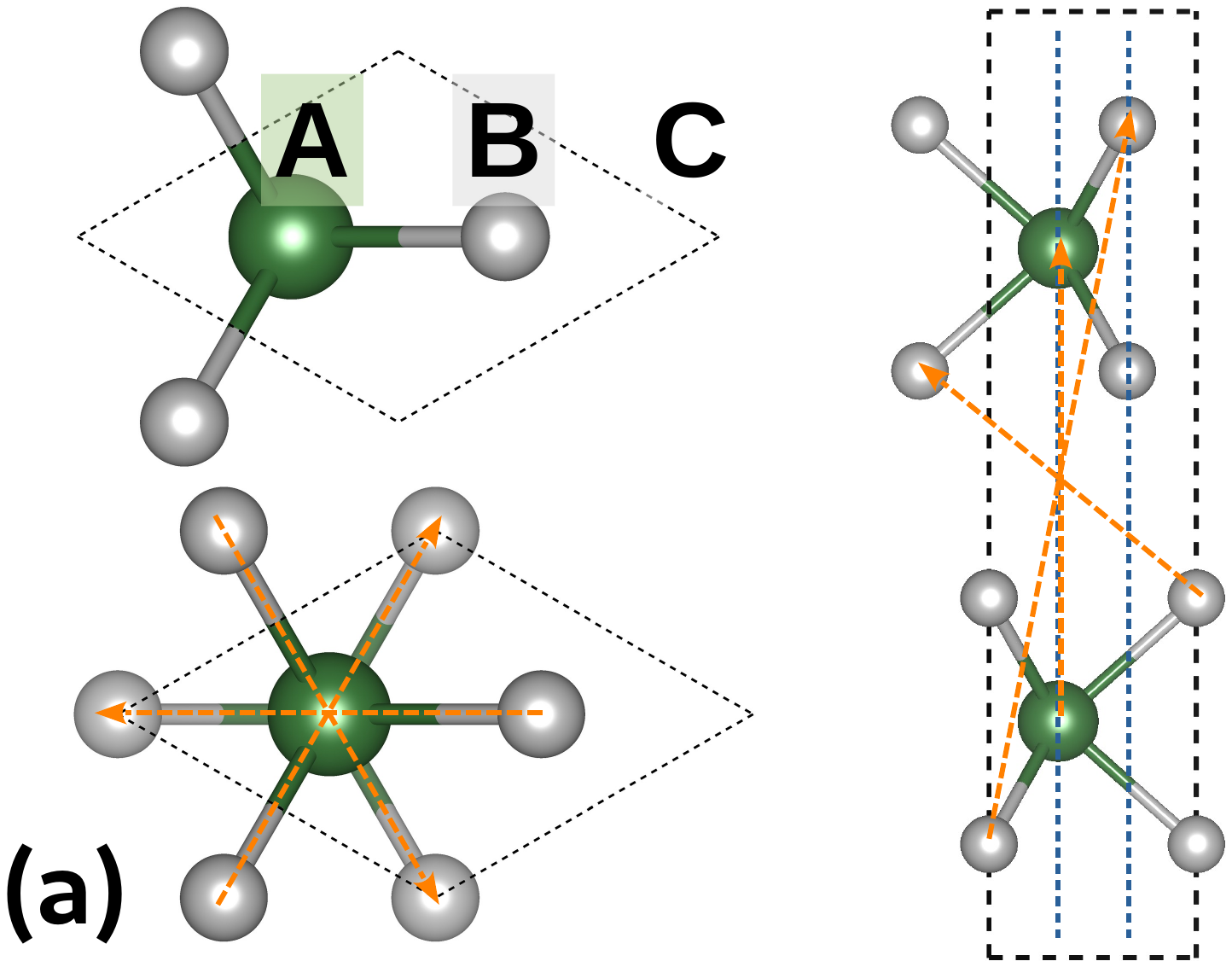}
\includegraphics[trim= 0cm 1.5cm 0cm 0cm, width=0.24\linewidth]{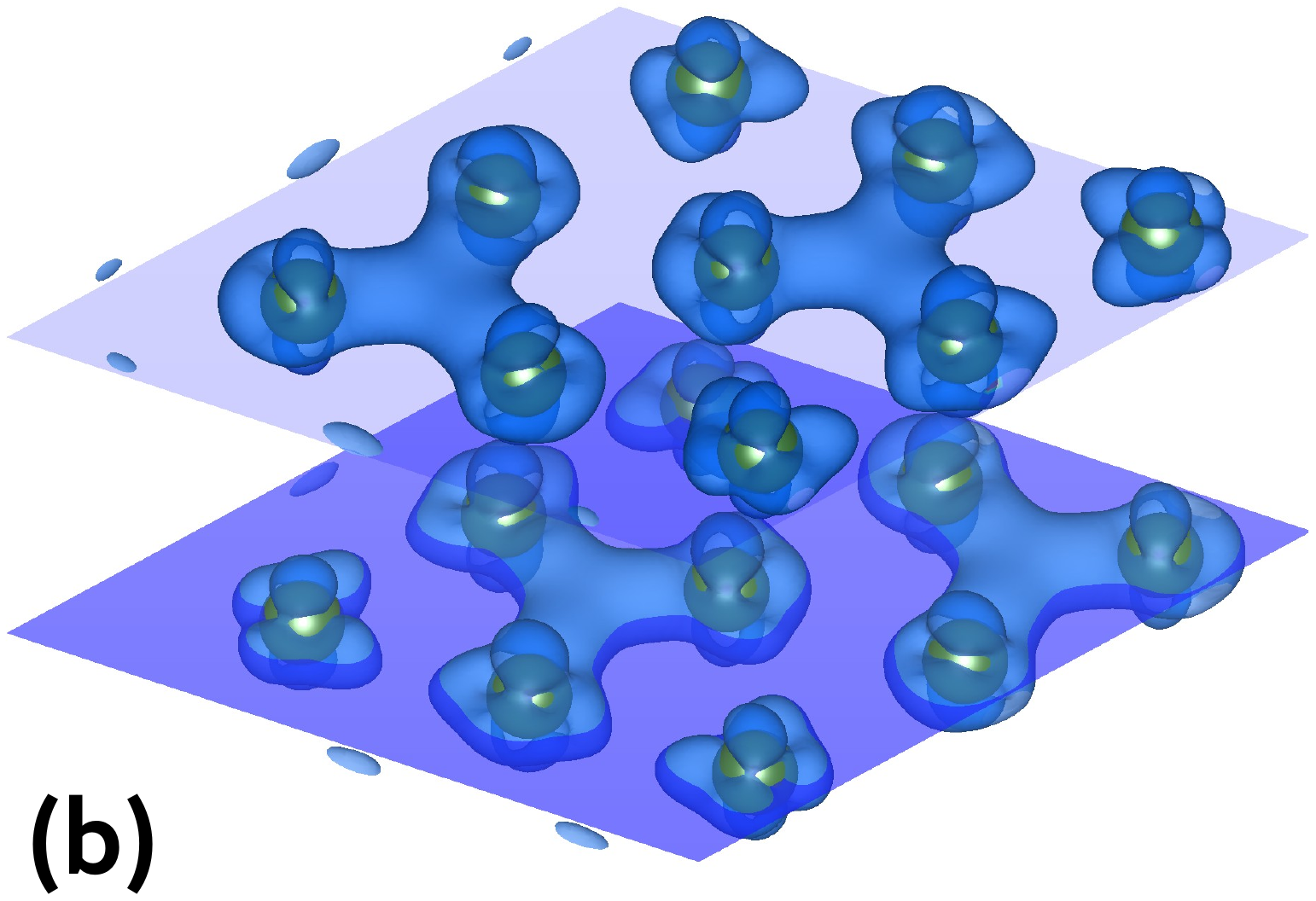}
\includegraphics[trim= 0cm 1.5cm 0cm 0cm, width=0.24\linewidth]{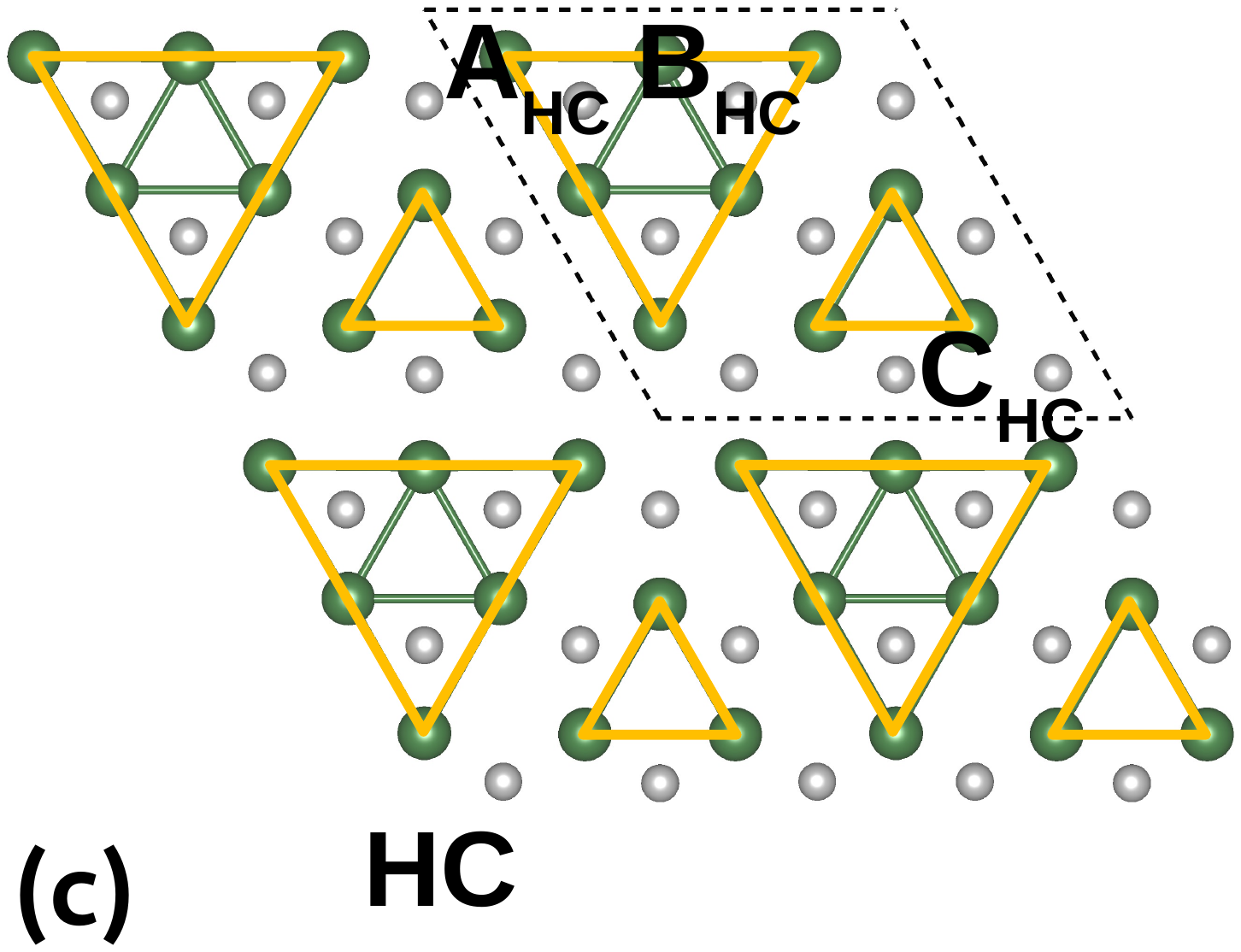}
\includegraphics[trim= 0cm 1.5cm 0cm 0cm, width=0.24\linewidth]{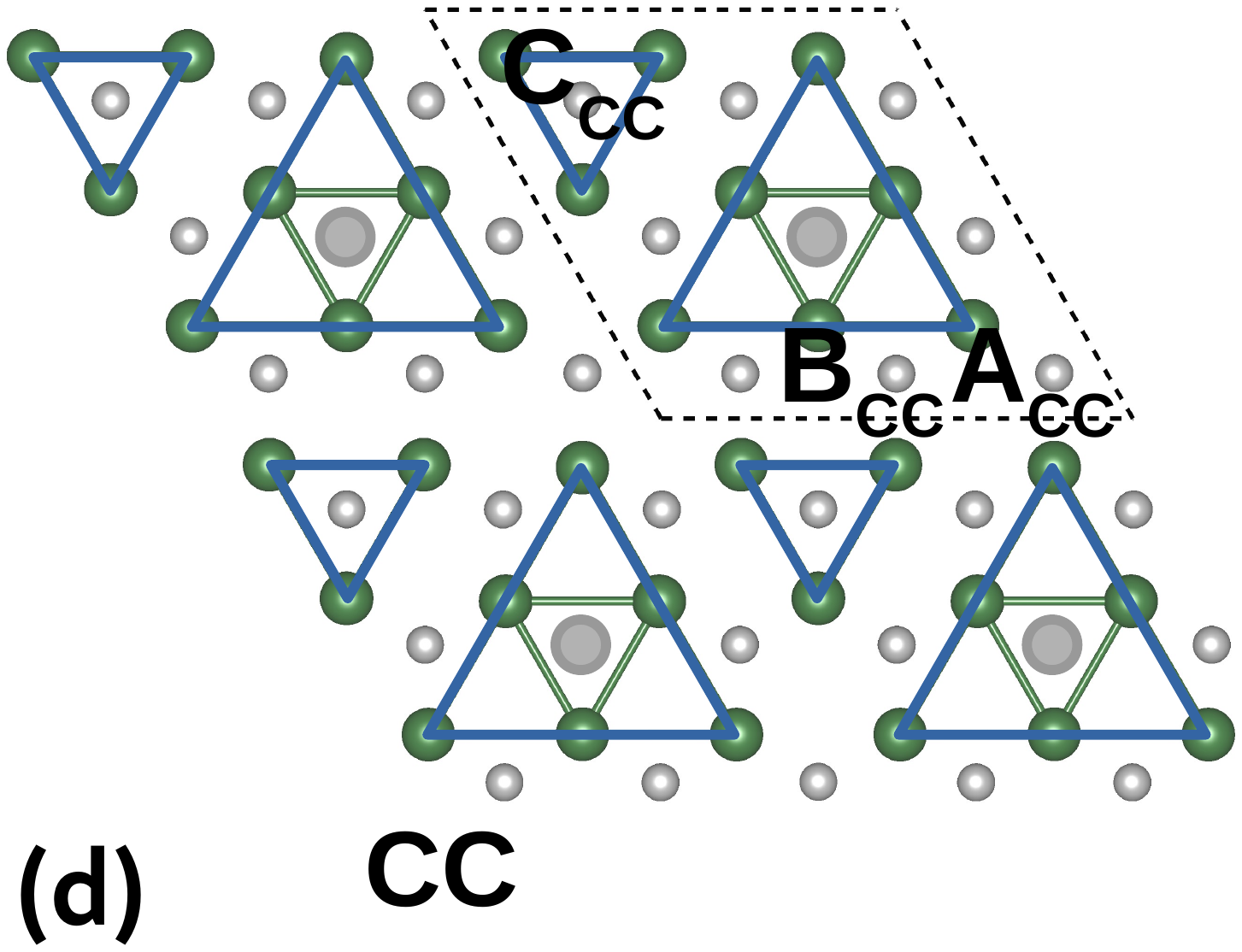}\\
\includegraphics[trim= 0cm 1.5cm 0cm 0cm, width=0.24\linewidth]{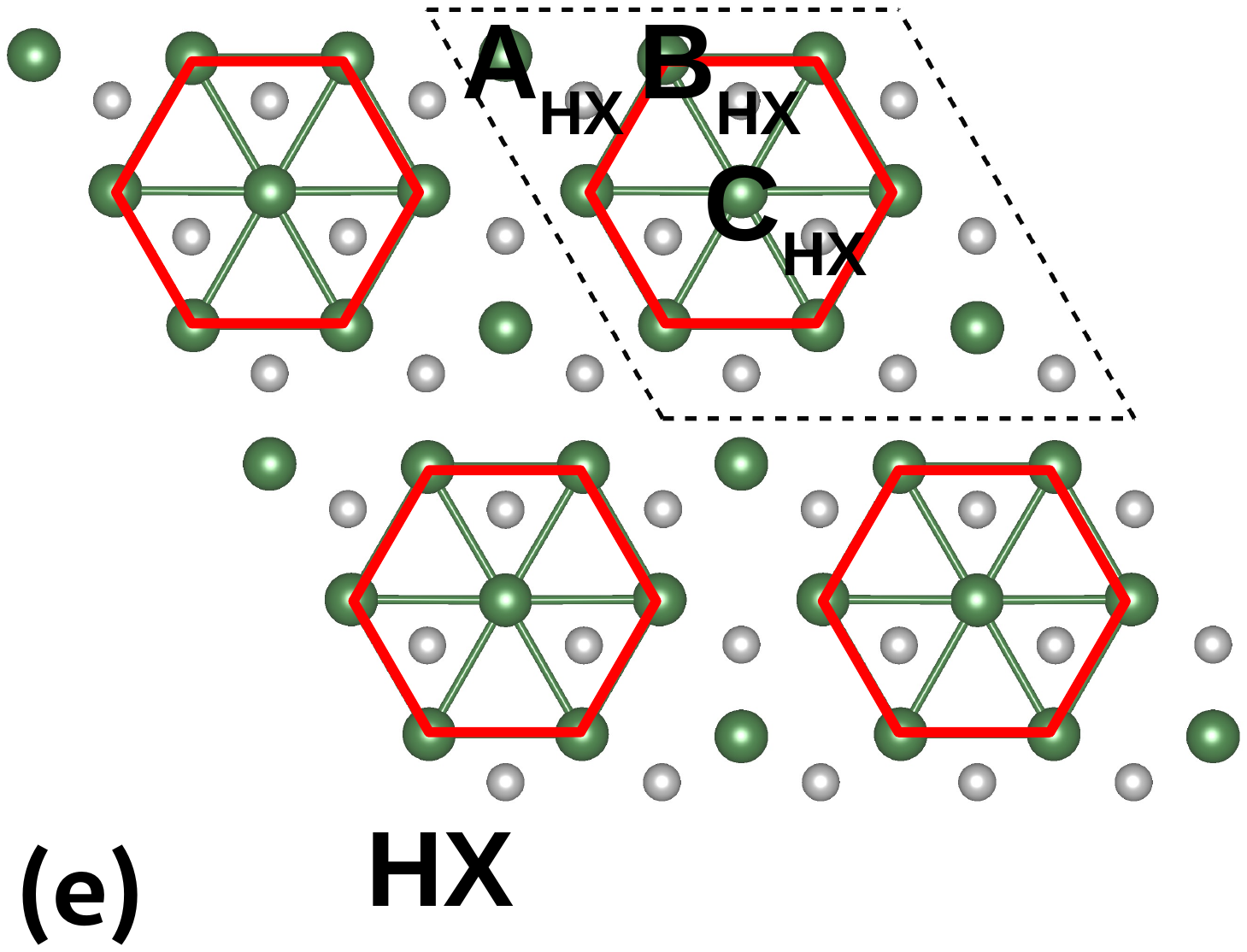}
\includegraphics[trim= 0cm 1.5cm 0cm 0cm, width=0.24\linewidth]{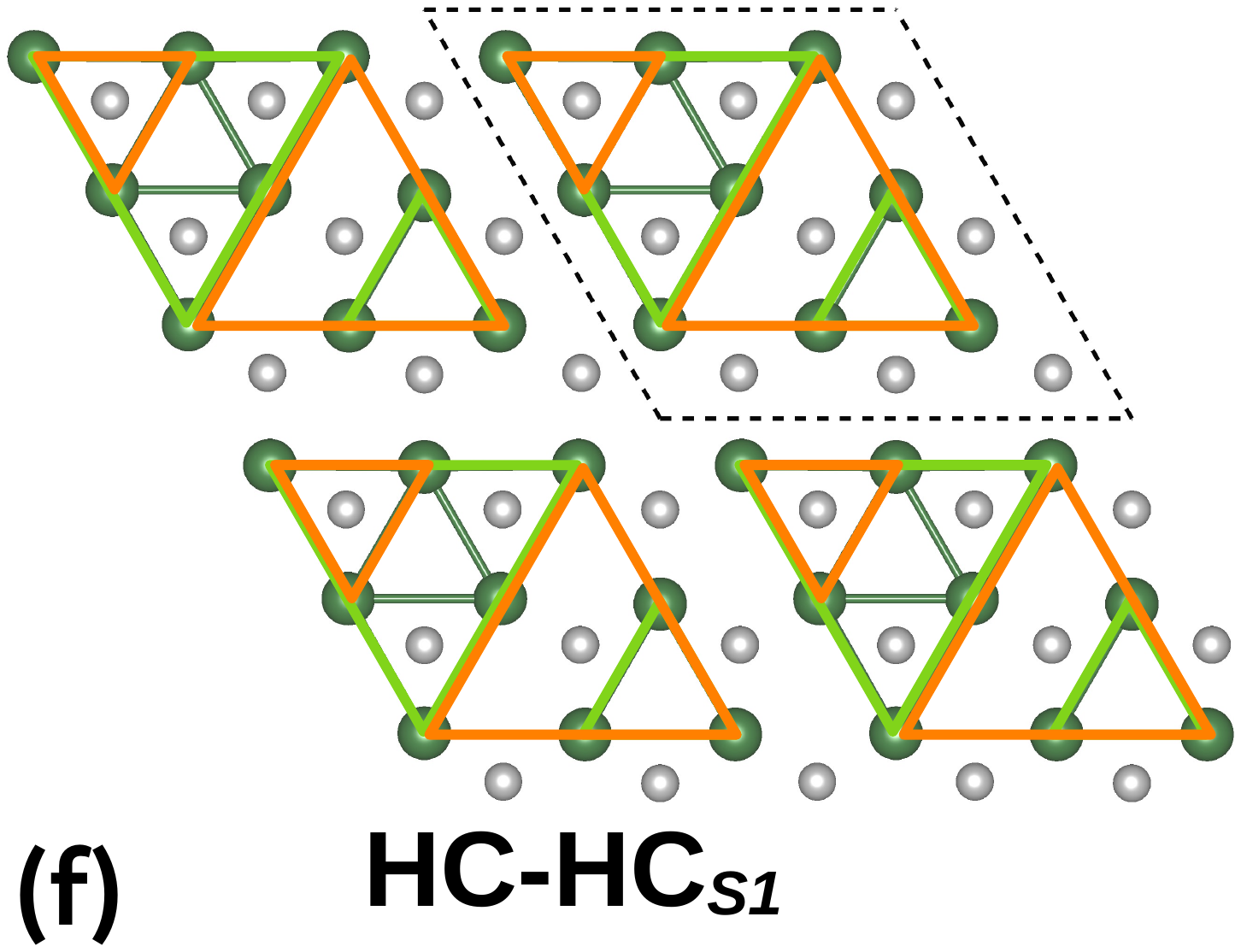}
\includegraphics[trim= 0cm 1.5cm 0cm 0cm, width=0.24\linewidth]{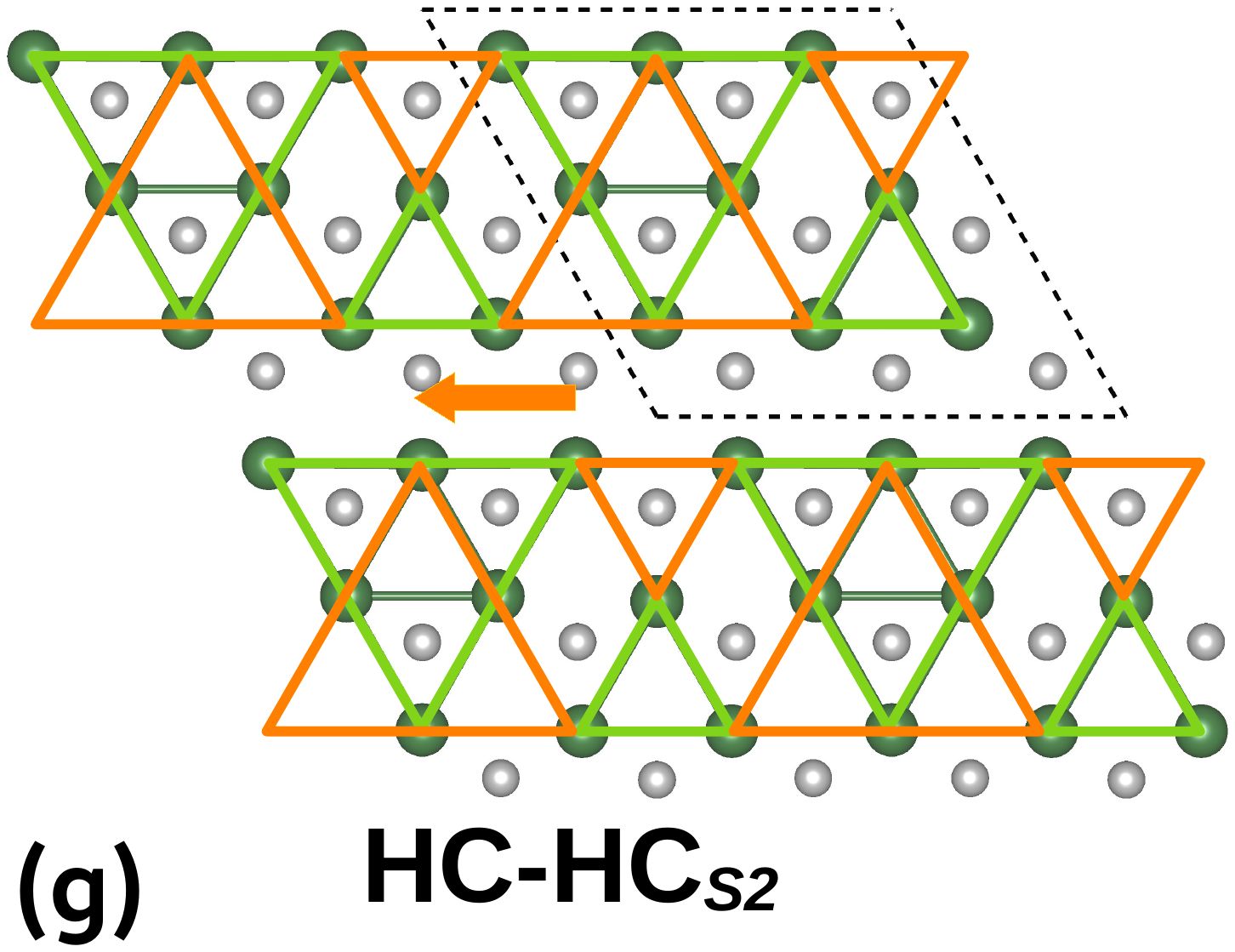}
\includegraphics[trim= 0cm 1.5cm 0cm 0cm, width=0.24\linewidth]{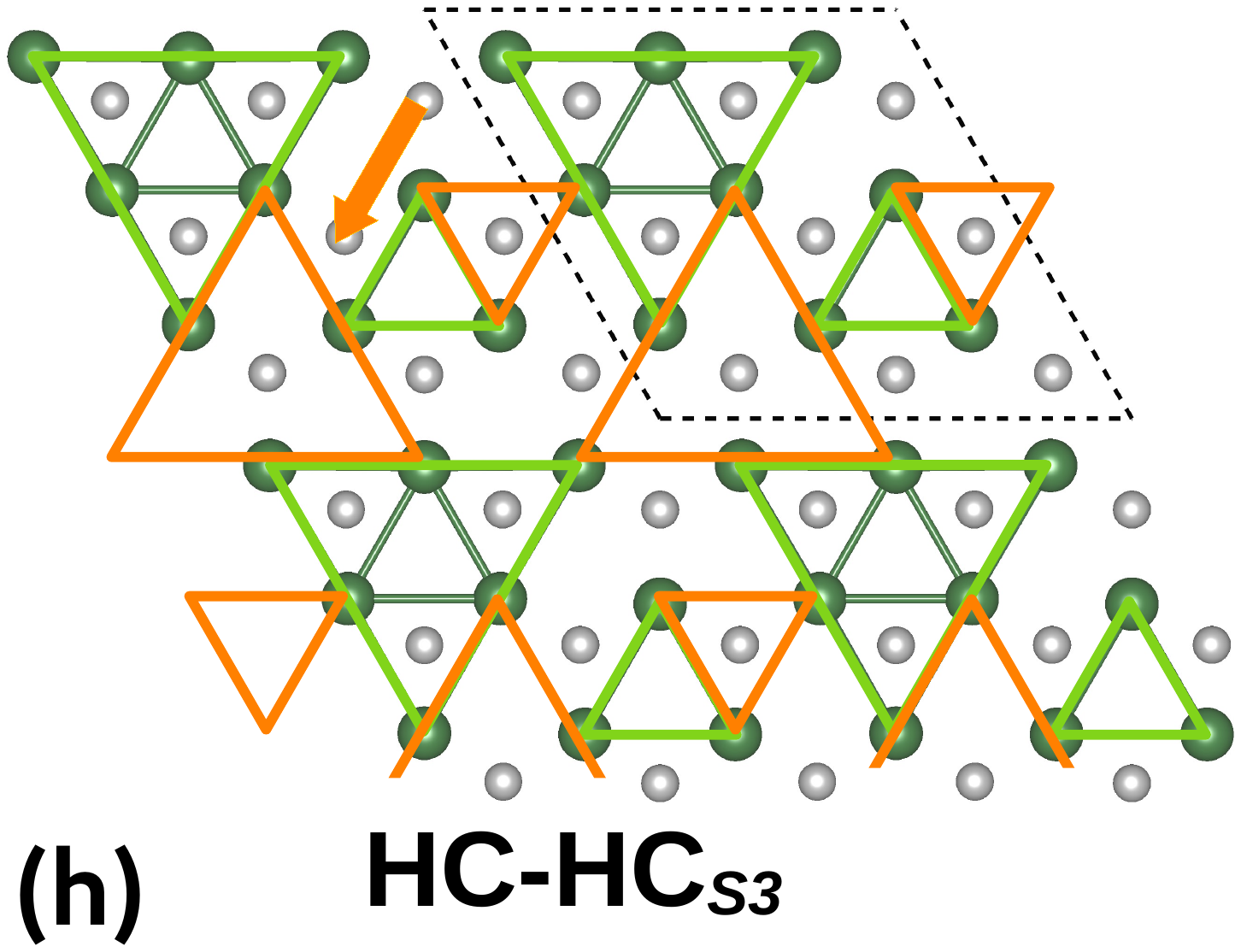}\\
\includegraphics[trim= 0cm 1.5cm 0cm 0cm, width=0.24\linewidth]{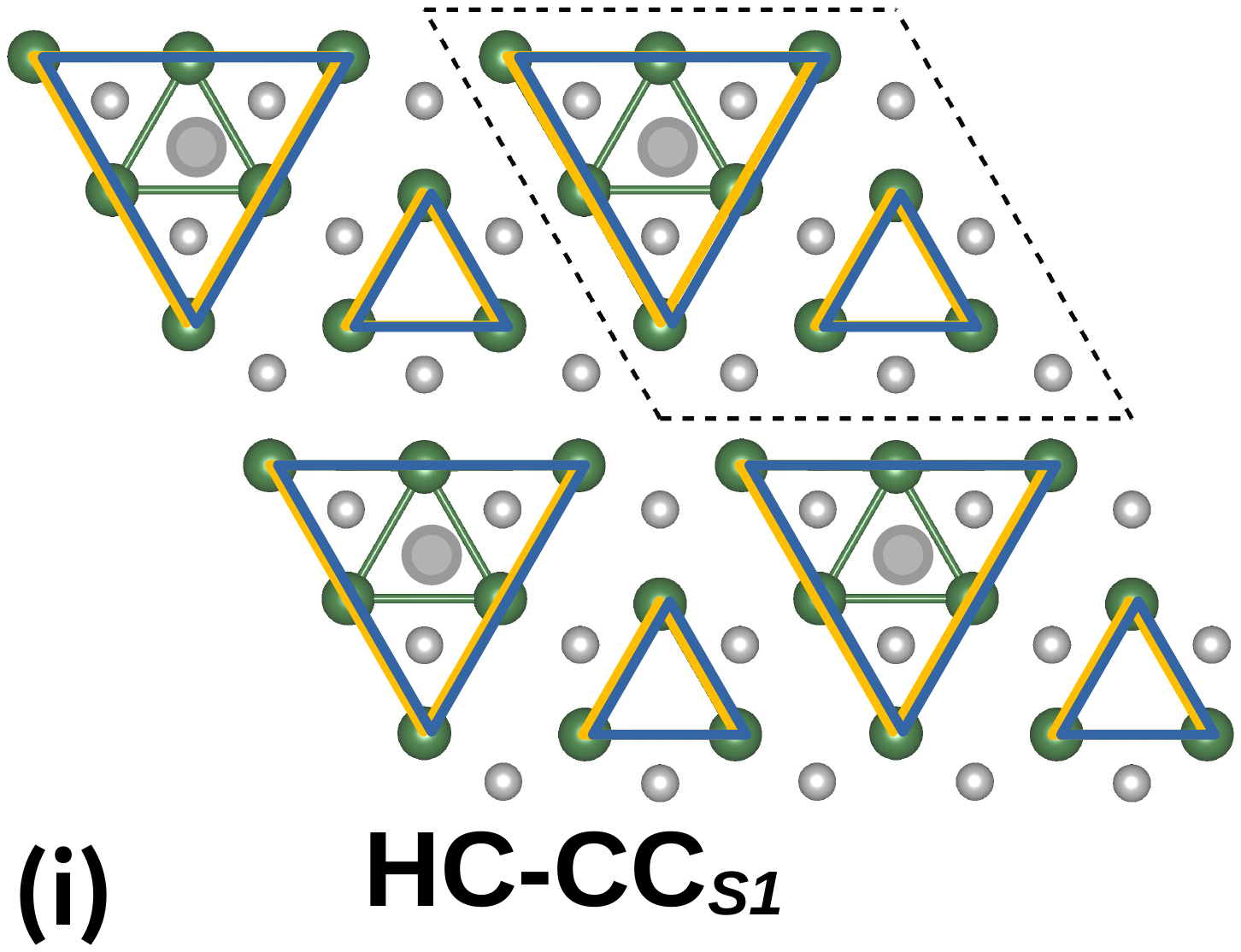}
\includegraphics[trim= 0cm 1.5cm 0cm 0cm, width=0.24\linewidth]{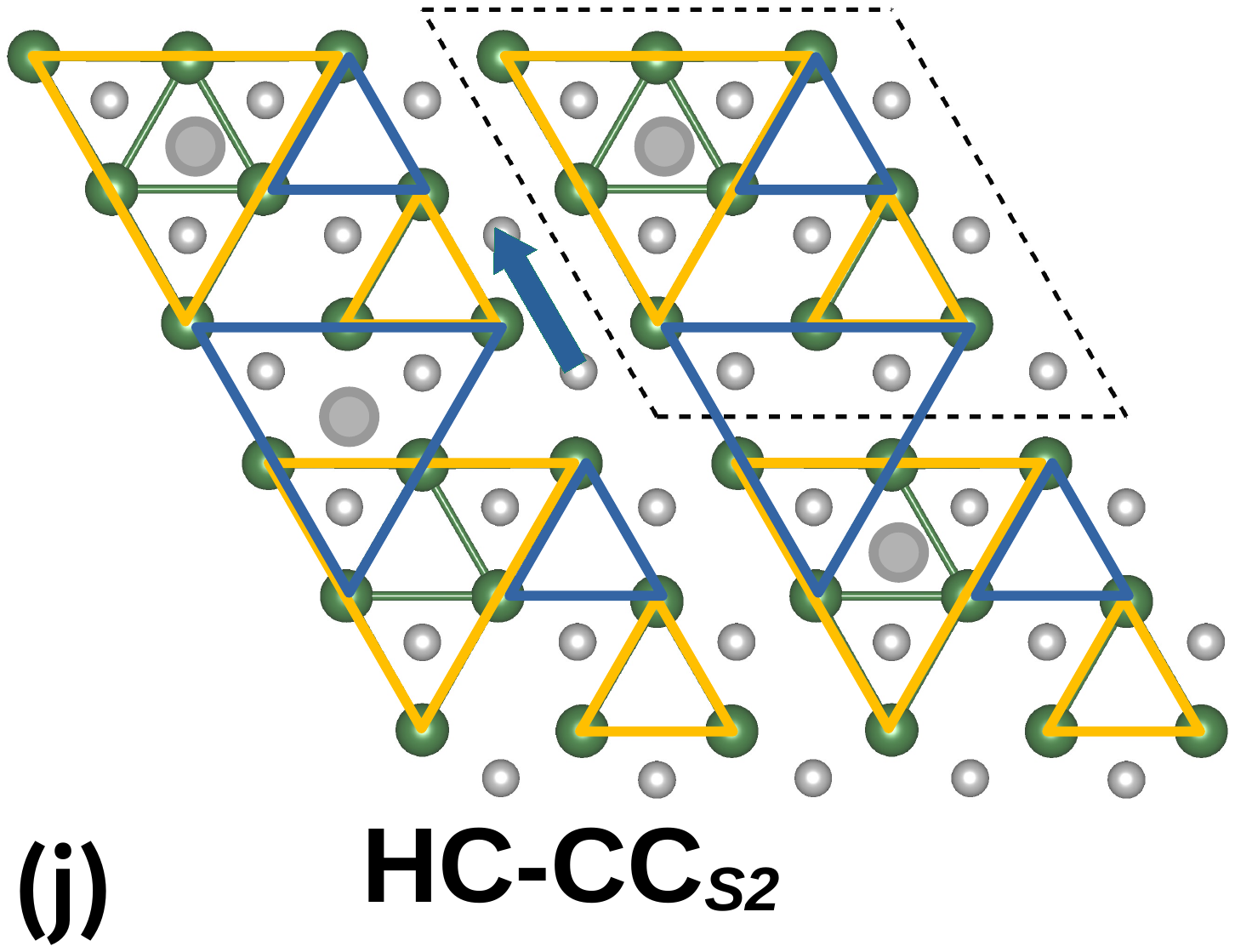}
\includegraphics[trim= 0cm 1.5cm 0cm 0cm, width=0.24\linewidth]{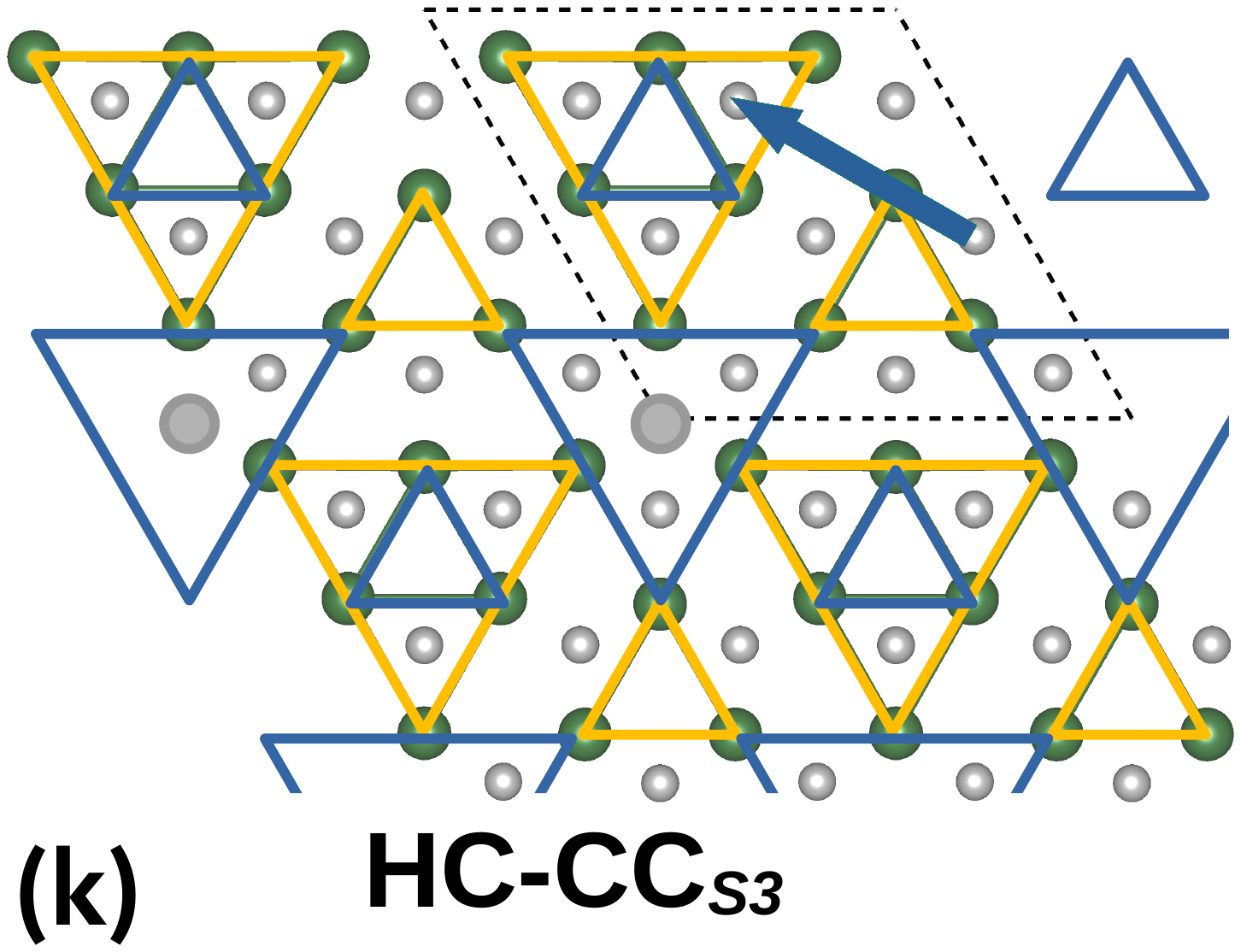}
\includegraphics[trim= 0cm 1.5cm 0cm 0cm, width=0.24\linewidth]{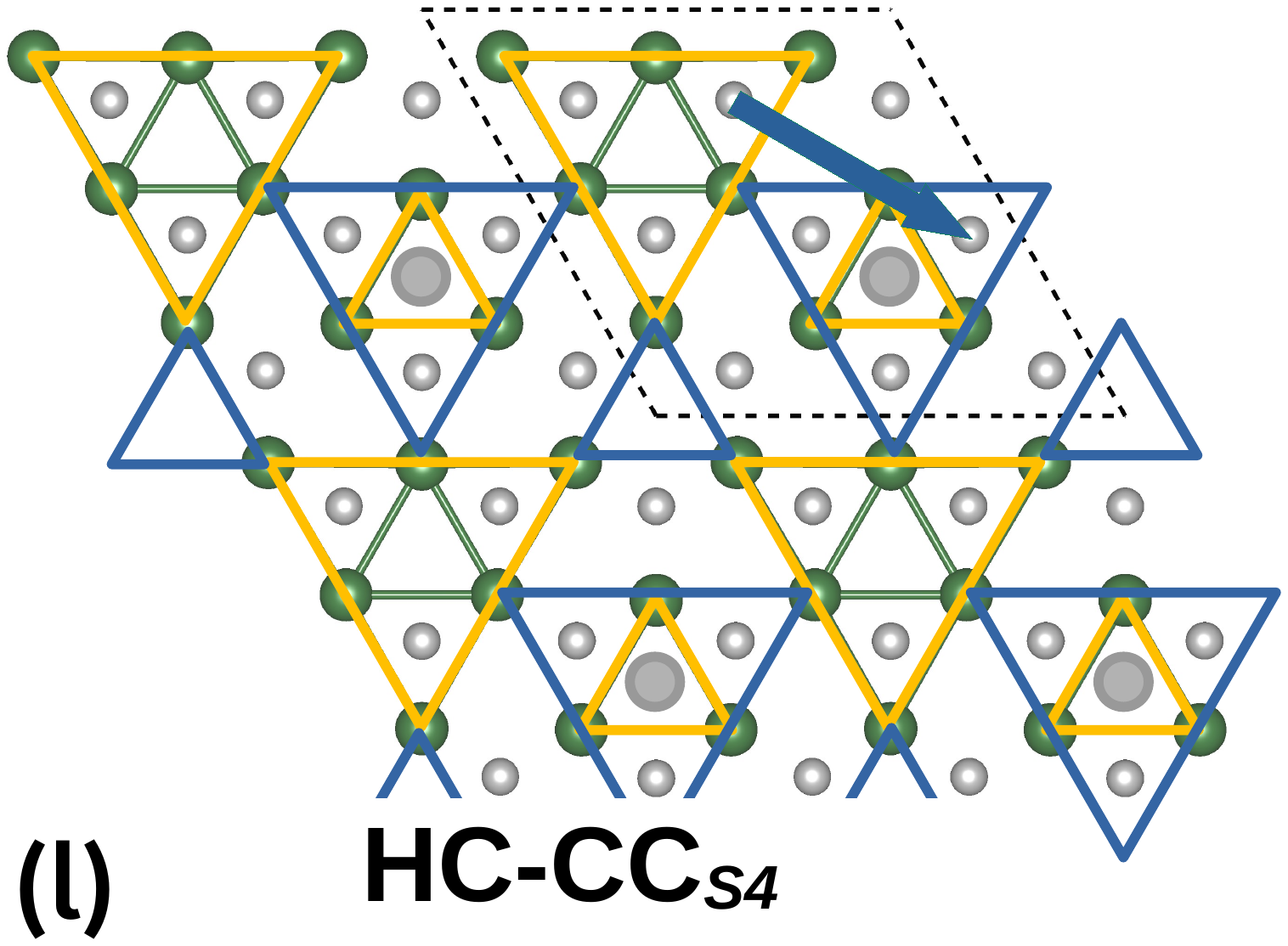}\\
	\caption{(a) Top view of the unit cell of 1H-NbSe$_2$ monolayer (top) and 2H-NbSe$_2$ bilayer (bottom), where Nb occupies the A site and Se the B site; on the right, side view of the unit cell for the stacking BAB-CAC of the 2H-type, showing the symmetry connecting the Se atoms. (b) Example of a CDW charge distribution across two composing monolayers in the 2H-NbSe$_2$ bilayer; the two planes are in correspondence via an inversion symmetry centered in the crossing point of the orange lines in panel (a). (c) HC, (d) CC and (e) HX structure of CDW in the 1H-NbSe$_2$ monolayer, with three nonequivalent points for each system. (f)-(h) The HC--HC blend in the 2H-NbSe$_2$ bilayer, with all possible in-plane displacements, obtained by a translation of a lattice vector of the orange lattice over the green one, which is depicted with arrows. (i)-(l) Ditto for the HC--CC blend. Note that the depiction of the displacement is arbitrary, since the two monolayers are not identical repetition of each other. For the HC--HC blend, we assume that a null displacement corresponds to the case when the rotation of the adjacent layer is performed around the Nb at the centre of the $3 \times 3$ supercell, as in panel (b). Instead, for the HC--CC blend, a null displacement corresponds to having A$_{CC}$, B$_{CC}$ and C$_{CC}$ overlap with A$_{HC}$, B$_{HC}$ and C$_{HC}$, respectively, as referred to panels (c) and (d).}
\label{fig:blend-displcm_summary}
\end{figure*}

Besides a few exceptions like 1T-TiSe$_2$, the CDWs in bulk TMDs have been traditionally assumed to retain a two-dimensional periodicity, which seems to find a plausible justification in the weak nature of the inter-layer coupling~\cite{Rossnagel_2011}.
Nevertheless, the inter-layer stacking was recently found to be 
sufficient for explaining the insulating character of bulk 1T-TaS$_2$ even without introducing an explicit Hubbard-type Coulomb correction~\cite{darancet_PRB_2014,ritschel_NP_2015,ritschel_PRB_2018,lee_PRL_2019,park_QM_2021,jung_PRB_2022,sayers_PRL_2023}. These works motivated similar research for other materials where effects related to a metal-to-insulator transition are expected to emerge in thin films, as e.g. 1T-TaSe$_2$~\cite{ji_PRB_2020,wang_AFM_2023}, 1T-NbS$_2$~\cite{wang_PRB_2022} and 1T-NbSe$_2$~\cite{dai_AFM_2023}.
As one may notice, all the aforementioned studies are focused on the role of the inter-layer stacking in structures of 1T-type, where a change in the conductive character is expected to leave fingerprints that are (relatively) easy to detect. One may naturally ask if an inter-layer stacking of CDWs emerges also in metallic TMDs and if this order can be detected through current experimental techniques. These questions have not only a fundamental, speculative motivation, but are also intimately connected to our ability to model the CDW state in solids, thin films, and monolayers. The aim of the present manuscript is precisely to shed light on these issues, by focusing on a bilayer of 2H-NbSe$_2$. 

In the last decade, 2H-NbSe$_2$ has had a central place in the study of CDWs in low-dimensional systems~\cite{xu_zq-NanoT2021}. 
Interest began when the strength of CDWs in thin films was discovered to increase for decreasing thickness, reaching its maximum value for the monolayer~\cite{frindt-PRL.28.299,xi_xx-nnano2015}. Subsequent studies have emphasized the crucial role played by substrate and sample preparation, while also pointing out problems in conciliating experimental results with our current theoretical understanding~\cite{ugeda-nphys2016,lin_dj-ncomm2020,bianco_PRL_2020, dreher_ACSNANO_2021,lin_NL_2022}.
The bulk stacking of the 2H-type, depicted in Figure~\ref{fig:blend-displcm_summary}(a), consists of bilayer units ordered as BAB--CAC, where A is a transition metal atom and B and C are chalcogen  atoms.
A bilayer is thus the building unit of the bulk, as well as the thinnest system keeping its centro-symmetric nature~\cite{myself-NPGAM2020}. The symmetry of the structure is a key factor in the discussion of subtle electronic effects, as it affects the number and character of the bands crossing the Fermi level, with important consequences on the direct-space modulation of CDWs as well as the gap opening associated to them~\cite{pasztor-ncomm2021,pasztor-PRR.1.033114}. 
This is also evident in the stability of thin films with respect to their thickness, which highlights the role played by bilayers in 2H-NbSe$_2$~\cite{darshana-preprint2023,samuely_PRB_2023}.

Despite these fundamental considerations, surface characterization experiments of 2H-NbSe$_2$, based on scanning tunnelling microscopy (STM) and transmission electron microscopy (TEM), are almost always modeled through electronic structure calculations of a monolayer~\cite{gye_gc-PRL.122.016403,myself-NPGAM2020,park-ncomm2019}. Although justified by the surface sensitivity of the aforementioned experiments as well as by computational convenience, this limitation has to be addressed to improve on our current theories, as already recognized in Refs.~\onlinecite{myself-NPGAM2020,oh_es-PRL.125.036804,zheng_NL_2022}. 

In the present manuscript, we investigate the electronic and structural properties of a bilayer of 2H-NbSe$_2$ through  density-functional theory (DFT). The inter-layer stacking of CDWs is analyzed in detail, with and without the inclusion of the vdW interaction. The analysis of the ground state and excited states shows that combining layers with CDWs of different type tends to be energetically unfavourable, but less than what expected from the data obtained for the 1H-NbSe$_2$ monolayer. Signatures of the inter-layer stacking are identified in the computed charge distributions, their Fourier transforms, as well as in the corresponding STM maps. This analysis not only clarifies the role of the inter-layer stacking in metallic system, but also provides a concrete guide for future experimental work on this subject or 2H-NbSe$_2$ in general.

\section{Methods}
 Electronic structure calculations are performed in DFT by means of the projected augmented wave (PAW) method, as implemented in the Vienna Ab-initio Simulation Package (VASP).
 Pseudopotentials are used to treat $4s^{2}4p^{6}5s^{1}5d^{4}$ for Nb and $4s^{2}4p^{4}$ for Se as valence electrons.
 The exchange-correlation functional is treated in the generalized gradient approximation (GGA), as parameterized by Perdew-Burke-Ernzerhof (PBE)~\cite{blochl-PRB.50.17953,kresse-PRB.59.1758}. The plane-wave basis is constructed with a cutoff energy of \SI{500}{\electronvolt}.
 Structural relaxation is performed using a quasi-Newton algorithm to minimise forces down to \SI{e-3}{\electronvolt/\angstrom} on each atom, while an energy tolerance of \SI{e-7}{\electronvolt} is imposed for the convergence of the electronic loop.
 For tasks requiring a more accurate determination of the electronic properties, such as plotting charge density distributions and density of states (DOS), the electronic loop is converged until the energy variation became lower than \SI{e-9}{\electronvolt} for the primitive unit cells and \SI{e-7}{\electronvolt} for the supercells.

 Calculations for the 2H-NbSe$_2$ bilayer are performed with a vacuum region of about 20~{\AA}, added along the $\hat{z}$. direction. The normal phase is modeled with the primitive unit cell, where the two composing monolayers form two sublattices whose in-plane coordinates are in correspondence to each other by an inversion symmetry, see Figure~\ref{fig:blend-displcm_summary}(a). In the 2H-NbSe$_2$ bulk, this becomes equivalent to a rotation by ${2n\pi}/{3} + {\pi}/{3}$ around the Nb lattice site plus a translation along $\hat{z}$. The in-plane lattice constant is set to 3.45~{\AA}, which is usually used to model monolayers~\cite{lebegue-PRB.79.115409,silvaguillen-2DMat2016,zheng-PRB.97.081101,myself-PRB.98.195419,myself-NPGAM2020,Sarkar_2022,watanabe2022}, corresponding to the substrates often adopted for growing monolayers and thin films~\cite{fang-ScienceAdv2018,nakata-npj2DMA2018,oh_es-PRL.125.036804}.
 The analysis of the CDW state is performed using $3 \times 3$ supercells, which can accommodate the distortions associated to the phonon instabilities~\cite{myself-NPGAM2020}.
 Each CDW corresponds to a metastable state that can be obtained via DFT by changing the initial positions of the ions before structural relaxation.
 A {\bf{k}}-mesh of $45 \times 45 \times 1$ ($20 \times 20 \times 1$) is used for the primitive unit cell (supercell), in combination with a Gaussian smearing of \SI{0.01}{\electronvolt}. The employed computational parameters have been determined to ensure that the energy differences reported in Table~\ref{cdw-enhi:tab} are well converged.

 The vdW interaction is taken into account within DFT, adding corrections to the PBE functional. As explained below, the normal phase is studied via several approaches, namely Grimme-D2 (GGA+D2)~\cite{grimme_s-JCompChem2006}, Tkatchenko-Scheffler (GGA+TS)~\cite{tkatchenko_a-PRL.102.073005}, non-local correction by M. Dion {\it{et al.}}~\cite{dion_m-PRL.92.246401} in the formulation of Klime\v{s} (GGA+DF)~\cite{klimes_j-JPCM2009} and many-body dispersion with  fractional ion (GGA+MBD@FI)~\cite{gould-PCCP2016,rehak-PCCP2020}.
 The supercell calculations for the CDW are performed by means of GGA+DF and GGA+TS. For the latter, we obtain results consistent with GGA+DF regarding the HC-HC and CC-CC blends, which supports our results. However, we did not manage to converge the mixed blends in CDW states with the correct symmetry. This instability is a consequence of the large reduction of the inter-layer distance, which will be discussed in the analysis of Table~\ref{vdw-str:tab}, where ${d}_{\text{Se-Se}}$ is shown to be reduced by more than 20\% with respect to the pure GGA value.
 These aspects are consistent with the results reported for the bulk~\cite{bucko_t-PRB.87.064110} and can be traced back to the general tendency of GGA+TS to overestimate the vdW bonding in some systems~\cite{Maurer_2019_review}.

 Finally, the STM simulations are performed by means of the BSKAN code~\cite{hofer-PSS2003,palotas-JPCM2005}, using the electronic structure obtained from the VASP calculations. The revised Chen method~\cite{mandi-PRB.91.165406} with an electronically flat and spatially spherical tip orbital is used, which is equivalent to the Tersoff-Hamann model of electron tunnelling~\cite{tersoff-PRB.31.805}.
 The reported STM images are in constant current mode obtained at -0.2~V bias voltage applied to the STM tip (with the maxima of the current contours at 5.8 \AA\ for all structures for  a better comparability). The visualization of the ionic positions and the charge density as well as the calculation of the structure factors are produced with VESTA JP-Minerals~\cite{VESTA-JACr2011}, supplemented by our in-house developed post-processing code. For completeness, the data before post-processing are reported in Appendix~\ref{sec:rawdata}. The analysis of the electronic properties is performed with the aid of VASPKIT~\cite{VASPKITnew}.

\section{Results}
\subsection{Structural properties in the normal phase}
The treatment of the vdW interaction affects significantly the inter-atomic distances in TMDs, inside a monolayer as well as across different monolayers~\cite{Bjorkman_2012,Twafik_2018}.
For bulk 2H-NbSe$_2$, this is well illustrated in the work by Bu{\v{c}}ko {\it{et al.}}~\cite{bucko_t-PRB.87.064110}, where the nearest neighbor distance $d_{\text{Nb-Nb}}$ between two Nb atoms belonging to different layers  is found to vary from 6.89~\AA, if vdW corrections are ignored, to 6.38~{\AA} and even 6.03~{\AA}, if vdW corrections are applied, via different approaches. The experimentally reported value, which corresponds to half the lattice parameter $c$, is around 6.27~\AA~\cite{MKRTCHYAN2018249,Naik_jpcc_2022}. The lack of an ideal method to account for the vdW interaction in NbSe$_2$ is further accentuated in systems composed of a few layers, since one needs to account for large regions of vacuum accurately~\cite{Twafik_2018}.
In fact, previous works on monolayer 1H-NbSe$_2$ have all used different DFT functionals for including vdW corrections~\cite{lian-NanoL2018.5,oh_es-PRL.125.036804,zheng_NL_2022}.
Owing to this uncertainty, before delving into the details of the CDWs, it is instructive to analyze how the basic structural properties depend on the choice of the DFT functional. Structural optimization of bilayer 2H-NbSe$_2$ was performed by DFT in generalized gradient approximation (GGA), without and with vdW corrections, using the functionals GGA+D2~\cite{grimme_s-JCompChem2006}, GGA+TS~\cite{tkatchenko_a-PRL.102.073005}, GGA+DF~\cite{dion_m-PRL.92.246401,klimes_j-JPCM2009}, and GGA+MBD@FI~\cite{gould-PCCP2016,rehak-PCCP2020}.  
\begin{table}[t]
\centering
 \caption{Structural parameters of the 2H-NbSe$_2$ bilayer in the normal phase, as computed via DFT with different functionals, with or without vdW corrections (see  main text). The distances between nearest neighbor pairs of Nb atoms ($d_{\text{Nb-Nb}}$) and Se atoms ($d_{\text{Se-Se}}$) belonging to different monolayers are reported. In addition, the distances between nearest neighbor pairs of Se atoms ($\tilde{d}_{\text{Se-Se}}$) belonging to the same monolayer are reported, alongside the corresponding Se-Nb-Se angle.}
 \label{vdw-str:tab}
 \begin{tabular}{l||c|c|c|c}
	 &   $d_{\text{Nb-Nb}}$({\AA})   &   $d_{\text{Se-Se}}$({\AA})   & $\tilde{d}_{\text{Se-Se}}$(\AA) & $\widehat{\text{SeNbSe}}$($^\circ$) \\
 \hline\hline
  GGA        &   6.926   &   3.560   &  3.373  & 80.5        \\
  GGA+D2     &   6.381   &   3.034   &  3.367  & 80.4        \\
  GGA+TS     &   6.053   &   2.732   &  3.355  & 80.2        \\
  GGA+DF     &   6.527   &   3.141   &  3.401  & 81.0        \\
  GGA+MBD@FI &   6.178   &   2.844   &  3.356  & 80.2        
 \end{tabular}
\end{table}
Results of the structural optimization of the normal phase, including the aforementioned $d_{\text{Nb-Nb}}$ and the analogous $d_{\text{Se-Se}}$, are reported in Table~\ref{vdw-str:tab}. For both distances, the largest values are found for plain GGA, as expected. Corrections resulting from the treatment of the vdW interaction are smallest for GGA+DF and largest for GGA+TS, which is in agreement with the trend of the bulk data discussed above~\cite{bucko_t-PRB.87.064110}. Since there are no experimental data to use as a reference for determining the best functional to model bilayer 2H-NbSe$_2$, we perform two sets of calculations for the CDWs, using GGA+DF and GGA+TS respectively. 
These two methods do not only correspond to the largest variation of the inter-layer distances in Table~\ref{vdw-str:tab}, but also originate from unrelated derivations. As explained in the Method section, not all possible CDWs could be stabilized in the correct symmetry with GGA+TS, while the most important structures could be found to be in good agreement with GGA+DF. Therefore, for sake of simplicity, only the results for GGA+DF will be presented and discussed in the rest of the article.

\subsection{Vertical stacking of CDWs}
  A wealth of literature produced during the last decade has contributed to identify several competing CDW structures in monolayer NbSe$_2$~\cite{calandra-PRB.80.241108,silvaguillen-2DMat2016,lian-NanoL2018.5,zheng-PRB.97.081101,myself-PRB.98.195419,gye_gc-PRL.122.016403,guster-NanoL2019.5,myself-NPGAM2020,oh_es-PRL.125.036804}.  Among those, three structures have a significantly lower energy, namely the hexagonal (HX), chalcogen-centred triangular (CC) and hollow centred triangular (HC) structures, shown in Figure~\ref{fig:blend-displcm_summary}(c-e).
  As illustrated on the right side of Table~\ref{cdw-enhi:tab}, these structures are all within an energy range of about 1 meV (GGA) or 2 meV (GGA+DF) per formula unit (f.u.), in agreement with the most recent literature~\cite{myself-PRB.98.195419,gye_gc-PRL.122.016403,myself-NPGAM2020}.
  When the CDW state is formed in a bilayer, there are more degrees of freedom to take into account. Since the HX, CC and HC structures dominate the physical properties of the bulk as well~\cite{calandra-PRB.80.241108}, we expect that the stacking of two or more layers is not going to affect the in-plane modulation so much as to involve higher energy solutions, as e.g. the star phase reported by Guster {\it{et al.}}~\cite{guster-NanoL2019.5}. 
  Therefore, we proceed to investigate all possible stacking configurations allowed by the symmetry of the previous modulations, including the mixing of different structures in the two monolayers forming the bilayer. 
  For clarity, we are going to refer to the combinations of possible structures, e.g. HC--HC or HX--CC, as CDW blends for the bilayer. Like for the stacking of stars of David CDWs in 1T-TaS$_2$~\cite{ritschel_PRB_2018}, only a limited number of nonequivalent configurations are possible for any given blend, constructed by combining a rotation of the layers with a translation vector of the direct lattice.  For simplicity, we are going to call these different configurations as displacements and label them with a subscript $S_n$, where $n$ is an integer up to number of possibilities allowed by symmetry. Hence, the CDWs arising from different blends and displacements may be uniquely identified by labels like e.g. HC--HC$_{S_2}$ or HC--CC$_{S_4}$. It is important to stress that both blend and displacement affect that full symmetry of the crystal in the CDW state, reducing it with respect to the normal phase. In fact, the space group goes from $P\bar{3}m1$ (normal phase) to $C2/m$ (HC--HC blend, all displacements), $P3m1$ (HC--CC blend, with the exception of HC--CC$_{S_2}$ which maintains only the inversion symmetry, $C1$), $P_1$ (CC--CC$_{S_2}$) and $C_m$ (CC--CC blend, all other displacements).
  An important difference between bilayers of the widely studied 1T-type and the present 2H-type is that the two composing monolayers are rotated by $\pi$ with respect to each other, and therefore even in the case of non-mixed blends, as e.g. HC--HC, there is no in-plane displacement for which the two sublattices (structural and/or charge modulations) are mapped into each other by a translation along $\hat{z}$.\\

  \begin{figure*}
\centering
\includegraphics[trim= 0cm 0.0cm 0cm 0cm, width=0.9\linewidth]{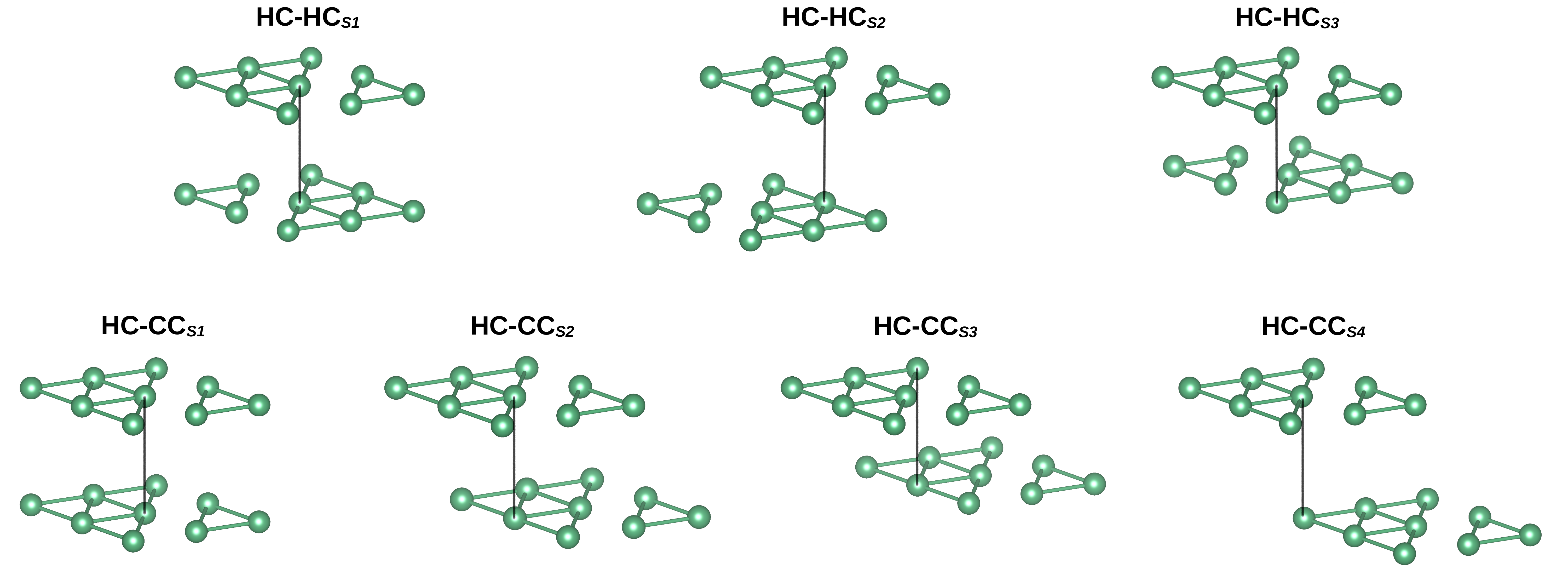}
	\caption{Axonometric view of the distribution of Nb atoms in the most relevant CDWs identified for the 2H-NbSe$_2$ bilayer, namely HC--HC$_{S_1}$, HC--HC$_{S_2}$, HC--HC$_{S_3}$, HC--CC$_{S_1}$, HC--CC$_{S_2}$, HC--CC$_{S_3}$, and HC--CC$_{S_4}$. The vertical black line indicates the alignment of the two monolayers. Although the HC--HC blend seems to correspond to Nb layers that only differ by a translation vector, the pattern of the Se atoms (not shown) breaks this symmetry when the two entire 1H-NbSe$_2$ monolayers are considered.}
\label{fig:sideviews}
\end{figure*}

The total energies of the optimised CDW structures in a bilayer of 2H-NbSe$_2$ are reported in Table~\ref{cdw-enhi:tab}. Blends not listed in the table possess a much higher energy and therefore are not discussed.
Independently from the inclusion of the vdW interaction, the most favorable energies are obtained for the HC--HC and HC--CC blends, whose geometries for all nonequivalent displacements are sketched in Figure~\ref{fig:blend-displcm_summary}(f-l).
For clarity, an alternative visualization is also reported in
Figure~\ref{fig:sideviews}.
In both GGA and GGA+DF, the ground-state of the bilayer NbSe$_2$ corresponds to HC--HC$_{S_3}$, shown in Figure~\ref{fig:blend-displcm_summary}(h).
This result is consistent with the STM data obtained for thin films \cite{oh_es-PRL.125.036804,ugeda-nphys2016}. The fact that the observed modulations are also in accordance with the STM data for the monolayer~\cite{gye_gc-PRL.122.016403,ugeda-nphys2016} suggests that the inter-layer coupling does not alter the fundamental character of the lattice instability. This is also confirmed by the same HC modulation being observed in bulk 2H-NbSe$_2$, via electron diffraction~\cite{wilson-AdvPhys2001}, and supported by angle-resolved photoemission spectroscopy (ARPES)~\cite{rahn-PRB.85.224532}. 
The first excited state is of the same blend as the ground state, namely HC--HC$_{S_1}$, shown in Figure~\ref{fig:blend-displcm_summary}(f). Instead, the second excited state is the most favourable configuration arising from a mixed blend, namely HC--CC$_{S_4}$, illustrated in Figure~\ref{fig:blend-displcm_summary}(l).

\begin{table}[b]
\centering
 \caption{Relative energies of the most competitive CDWs identified in the 2H-NbSe$_2$ bilayer, as computed via GGA and GGA+DF. Energies are given in meV/f.u. and with respect to the most favorable configuration, denoted as the ground state `GS'. CDWs from the CC--HX and HX--HX blends are not shown, since they have much higher energies. On the right side, corresponding results for the 1H-NbSe$_2$ monolayer are reported, alongside the closest bilayer group. The solution labelled as `sym' refers to the normal phase, without CDW, for both bilayer and monolayer.}
 \label{cdw-enhi:tab}
 \begin{tabular}{c|c|cc||c|cc}
    \multicolumn{4}{c}{bilayer} & \multicolumn{3}{c}{monolayer} \\
    \hline
		     &   &   GGA  &   GGA+DF  & & GGA & GGA+DF \\
 \hline\hline
 {sym} &  {} &  3.78   &   4.94   & {sym} &  3.94 &  6.11 \\
 \hline
	       HC--HC & $S_1$ &   0.06   &   0.21   & HC &   GS   &   GS   \\
		     & $S_2$ &   0.23   &   0.72   &    &        &        \\
		     & $S_3$ &    GS     &    GS     &    &        &        \\
 \hline                                         
	       HC--CC & $S_1$ &   0.31   &   0.82   & CC &  0.51 &  1.43 \\
		     & $S_2$ &   0.28   &   0.57   &    &        &        \\
		     & $S_3$ &   0.44   &   1.24   &    &        &        \\
		     & $S_4$ &   0.18   &   0.39   &    &        &        \\
 \hline                                         
	       HC--HX & $S_1$ &   1.02   &   1.34   & HX &  1.18 &  2.06 \\
		     & $S_2$ &   0.51   &   0.74   &    &        &        \\
		     & $S_3$ &   0.66   &   1.18   &    &        &        \\
 \hline                                         
	       CC--CC & $S_1$ &   0.57   &   1.11   &    &        &        \\
		     & $S_2$ &   0.48   &   1.03   &    &        &        \\
		     & $S_3$ &   0.59   &   1.36   &    &        &
 \end{tabular}
\end{table}

While the order of the lowest lying states does not change between GGA and GGA+DF, their relative energies depend on them. Table~\ref{cdw-enhi:tab} demonstrates that in both the monolayer 1H-NbSe$_2$ and bilayer 2H-NbSe$_2$ the stability of the CDWs increases when the vdW corrections are considered, as a consequence of the reduced inter-layer distances. The larger orbital overlap also increases the energy differences between the various blends and displacements, as expected. Interestingly, the mixed blends are not particularly disfavored by this mechanism, despite the reduced coherence of their chemical bonds.

Concerning the differences between monolayer and bilayer, Table~\ref{cdw-enhi:tab} shows that the stability of the CDW with respect to the normal phase is stronger in the former. This is consistent with existing observations on the trend of the critical temperature with respect to layer thickness~\cite{xi_xx-nnano2015,lin_dj-ncomm2020}, albeit there is still an on-going controversy regarding the role played by substrate and sample preparation~\cite{ugeda-nphys2016,lin_dj-ncomm2020,bianco_PRL_2020, dreher_ACSNANO_2021,lin_NL_2022}. 
Intuitively, the higher critical temperature observed in the bilayer can be rationalized in terms of inter-layer coupling, as the larger coherence of the Se-Se inter-layer bonds in the normal phase increases its stability with respect to CDW state, which has a larger periodicity~\cite{lin_dj-ncomm2020}.
Regarding the landscape of excited states, the energy difference between the HC and CC structures for the 1H-NbSe$_2$ monolayer amounts to 0.51 (1.43) meV/f.u. in GGA (GGA+DF). This value can be compared with the energy difference between the HC--HC$_{S_3}$ and CC--CC$_{S_2}$ structures for the 2H-NbSe$_2$ bilayer, which amounts to 0.48 (1.03) meV/f.u.
This reduction is less trivial to explain without implying a direct decrease of the instability associated to the electron-phonon coupling, in accordance with the analysis by Xi {\it{et al.}}~\cite{xi_xx-nnano2015}.
As a result of the weaker strength of the CDWs as well as the increased number of degrees of freedom, the energy landscape reported in Table~\ref{cdw-enhi:tab} shows that the excited states are much closer to the ground state in the 2H-NbSe$_2$ bilayer than in the 1H-NbSe$_2$ monolayer. 
An interesting question arising from these data is how they are going to evolve in a corresponding model for the bulk, where we expect a convergence behavior analogous to that exhibited by stacking fault energies for close-packed solids.

Focusing on the first three excited states identified above for the bilayer, we can see that their energies all lie within a range of 0.18 (0.39) meV/f.u. in GGA (GGA+DF). This energy range is smaller than the difference between HC and CC structures in the monolayer, which was shown to be overcome by defects, leading to domain formation observable by STM~\cite{oh_es-PRL.125.036804}.
Furthermore, the excitation energies are also much smaller than the experimental critical temperature, which suggests that thermal effects may also lead to a non-trivial mixing in the vertical stacking~\cite{lee_PRL_2019}. While a proper analysis of the stability with respect to thermal fluctuations would require the evaluation of the potential energy barriers between the various metastable states, an estimate can be obtained by comparing our results to existing modeling for 1T-TaS$_2$ bulk. For this system, thermally induced transitions between the two most favorable stacking configurations were demonstrated to be locked at about 60~K~\cite{lee_PRL_2019}. The energy difference between these states amounts to 0.08 meV/f.u. in GGA+TS, which is less than half what we obtain for the 2H-NbSe$_2$ bilayer in GGA+DF. Therefore, we may reasonably expect the stacking predicted here to be stable at temperatures below the experimental critical temperature. Another thermal effect worth considering is the correction due to the vibrational entropy, which has the potential to overcome the small energy differences discussed so far. Previous results for the symmetric 2H-NbSe$_2$ bilayer from GGA+D3 calculations showed that large translations (up to 3.5~{\AA}) of one monolayer with respect to the other one may induce a correction as large as 0.4 meV/f.u. at 10~K~\cite{acsnano.2c03157}. This correction is the consequence of the different degree of overlap between the orbitals of Se atoms belonging to different monolayers, see the right side of Fig.~\ref{fig:blend-displcm_summary}(a), which may change substantially the energy of the shear mode. For the CDWs investigated here, this strong variation of overlap is never possible, since the atomic displacements are small (up to 0.2~{\AA}) and not in phase. Thus, we expect a correction that is at least an order of magnitude smaller, which would not be sufficient to change the energy hierarchy of Table~\ref{cdw-enhi:tab}. Note that this scenario may be very different for bilayer structures of the 1T-type, where the composing monolayers may potentially generate very different shear modes depending on the vertical stacking of the CDWs. A more precise answer beyond this speculation would require the calculation of the phonon spectra of the bilayer with the CDWs with a sufficient accuracy to resolve these tiny contributions. This task would be outside the scope of the present manuscript.

In summary, these results show that the HC structure dominates the CDW landscape, while the CC structure is still close enough in energy to appear in STM images of bilayers, via mixed blends; the HX structure is too high in energy to be visualized in a perfect system, but it may probably arise near impurities, as previously reported for 1H-NbSe$_2$ monolayers~\cite{arguello-PRB.89.235115,liebhaber-NanoL2020.1,myself-PRB.98.195419}.


\subsection{Electronic properties}
\begin{figure}[t]
\centering
\includegraphics[trim= 0cm 0.0cm 0cm 0cm, width=0.92\linewidth]{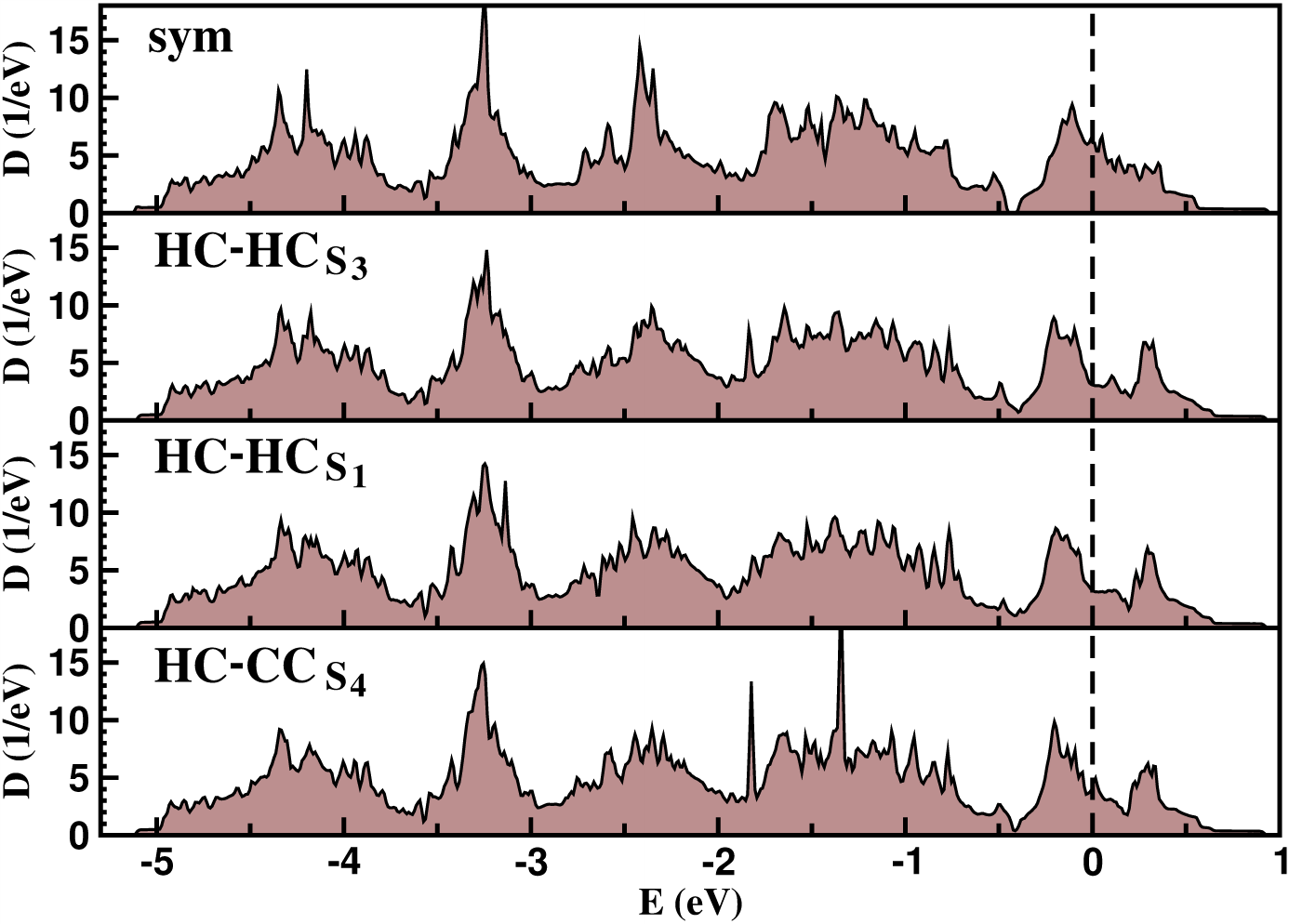}
\includegraphics[trim= 0cm 0.0cm 0cm 0cm, width=0.92\linewidth]{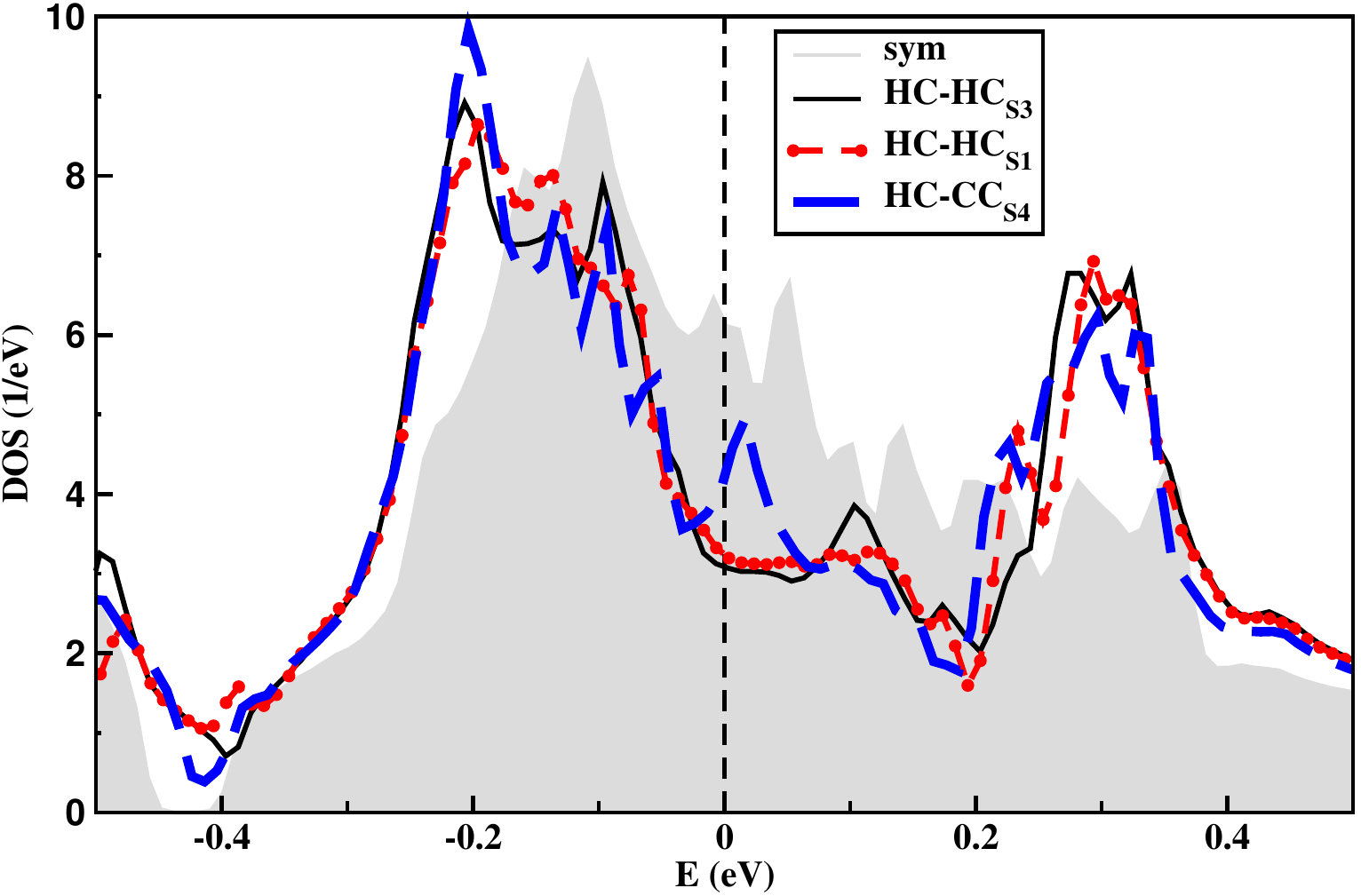}
	\caption{Top panel: total DOS of the CDW structures identified as the most energetically favorable in the 2H-NbSe$_2$ bilayer, compared to the symmetric state. The Fermi level is at zero energy. Bottom panel: magnified view around the Fermi level.}
\label{fig:TDOS}
\end{figure}
\begin{figure*}[t]
\centering
\includegraphics[trim= 0cm 0.0cm 0cm 0cm, width=0.95\linewidth]{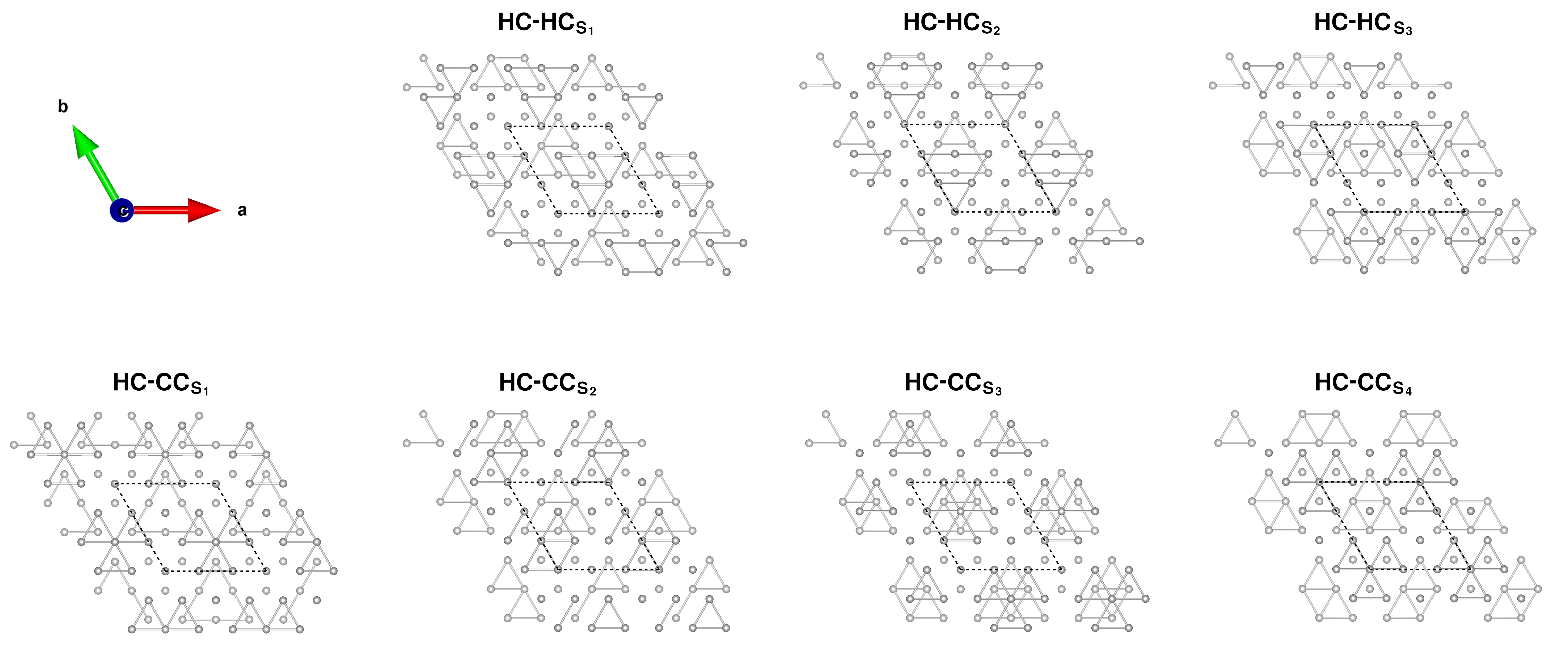}
	\caption{Se-Se bond distance patterns in the 2H-NbSe$_2$ bilayer for selected CDW structures, as calculated in GGA+DF.}
\label{fig:Sepatterns}
\end{figure*}
The total DOS of the 2H-NbSe$_2$ bilayer is illustrated in Figure~\ref{fig:TDOS}, for the symmetric state as well as for the most favourable CDW configurations. The general shapes of the four curves are very similar and the differences induced by the stacking across the whole valence band are minimal. An interesting feature is the formation of a few sharp peaks in certain configurations, as e.g. the peak at -1.8~eV for HC--CC$_{S_4}$. The analysis of the projected DOS for individual Se atoms (data not shown) makes it possible to identify that these sharp peaks originate from the $p_z$ orbitals of the Se atoms comprised between the two monolayers. This suggests that the periodic lattice distortion in each monolayer may lead to pairs of Se atoms in the out-of-plane direction that are characterized by an increased coherence, which in turn affects the energetic stability of the CDW configurations. This argument can be seen as a generalization of the mechanism proposed in Ref.~\onlinecite{lin_NL_2022} to explain the thickness dependence observed in Raman spectroscopy.
It is also instructing to analyze the distributions of states in the vicinity of the Fermi level, which is shown in the bottom panel of Figure~\ref{fig:TDOS}. A clear correlation is visible between the energy gained by the CDW structures and the depletion of states at the Fermi level. The same effect may also be observed among the different CDWs in the 1H-NbSe$_2$ monolayer, with HC showing a larger depletion than CC and HX~\cite{myself-PRB.98.195419}. 
The gain in (electronic band) energy associated to the depletion of states at the Fermi level has been largely investigated in literature, since it is a fundamental aspect of the one-dimensional Peierls model and its generalization to two- and three-dimensional systems, where it respectively induces a full gap or a pseudogap~\cite{Rossnagel_2011}. 2H-NbSe$_2$ is more complex in this regard, since the CDW originates from the momentum dependence of the electron-phonon coupling~\cite{PhysRevResearch.5.013218} and only a tiny fraction of electrons are involved in its formation~\cite{pasztor-ncomm2021,pasztor-PRR.1.033114}. Therefore, it is very interesting that our data show such a clear correlation between the energetic order and the depletion of states at the Fermi level.

From the total DOS reported in Figure~\ref{fig:TDOS}, we can infer that the transport properties exhibited by different stacking configurations of CDWs are going to be rather similar.
Lacking a major effect as the metal-insulator transition with a hysteresis cycle exhibited in 1T-TaS$_2$, one may wonder if there are other, more subtle consequences of the stacking configuration arising in 2H-NbSe$_2$, for the bilayer as well as for the bulk. 
Fundamentally, we are interested in identifying signatures of the CDW stacking in a metallic system that may be detected in possible experiments, being for a homogeneous, uniform structure, or for a coexistence of domains, in analogy to the work by Oh {\it{et al.}}~\cite{oh_es-PRL.125.036804}. In the following, we will explore two options, i.e. via STM measurements or structure factors analysis.

\subsection{Se patterns and STM simulations}

In the past years, comparison between theoretical and experimental STM images has given an important contribution to the research on monolayer, bulk and thin-films of 2H-NbSe$_2$~\cite{xu_zq-NanoT2021}.
Although STM experiments are by construction surface sensitive, one may think about the possibility of discriminating between stacking configurations by analysing their STM characteristics. The idea behind it is that STM is not only sensitive to the surface topography but also to the electronic structure, which is likely modulated in 2H-NbSe$_2$ depending on the variation of the coupled monolayer structures and the stabilized CDWs in each layer.
In STM, one has usually access to the Se-Se bond pattern of the top sublattice, which thus becomes of crucial relevance. The Se atoms display long/short bond modulations along the vertical direction as well as in the plane parallel to the Nb plane. 
These modulations accompany the Nb-Nb bond patterns~\cite{silvaguillen-2DMat2016,myself-PRB.98.195419}, as illustrated in Ref.~\onlinecite{myself-PRB.98.195419} for the 1H-NbSe$_2$ monolayer. 
In the latter, in-plane Se-Se bonds form triangular (hexagonal) patterns around the HC layer when coupled to HC and CC (HX); they form hexagonal patterns around the CC layer when coupled to the HC; finally, they form triangular patterns around the HX layer.
In the 2H-NbSe$_2$ bilayer, the Se-Se patterns depend on the CDW blend and displacement and are therefore a secondary effect of the inter-layer coupling that stabilizes these states.
The patterns of the Se-Se bond distances in the most relevant CDW states are illustrated in Figure~\ref{fig:Sepatterns}, as calculated in GGA+DF. We notice that, for each blend, the lowest energy corresponds to Se-Se patches which do not superimpose, see HC--HC$_{S_3}$ and HC--CC$_{S_4}$. One may speculate that what makes the HC--HC$_{S_1}$ and HC--HC$_{S_2}$ still relatively low in energy may be a minimal overlap of the neighbouring Se-Se patterns.
\begin{figure*}
\centering
\includegraphics[width=1.0\linewidth]{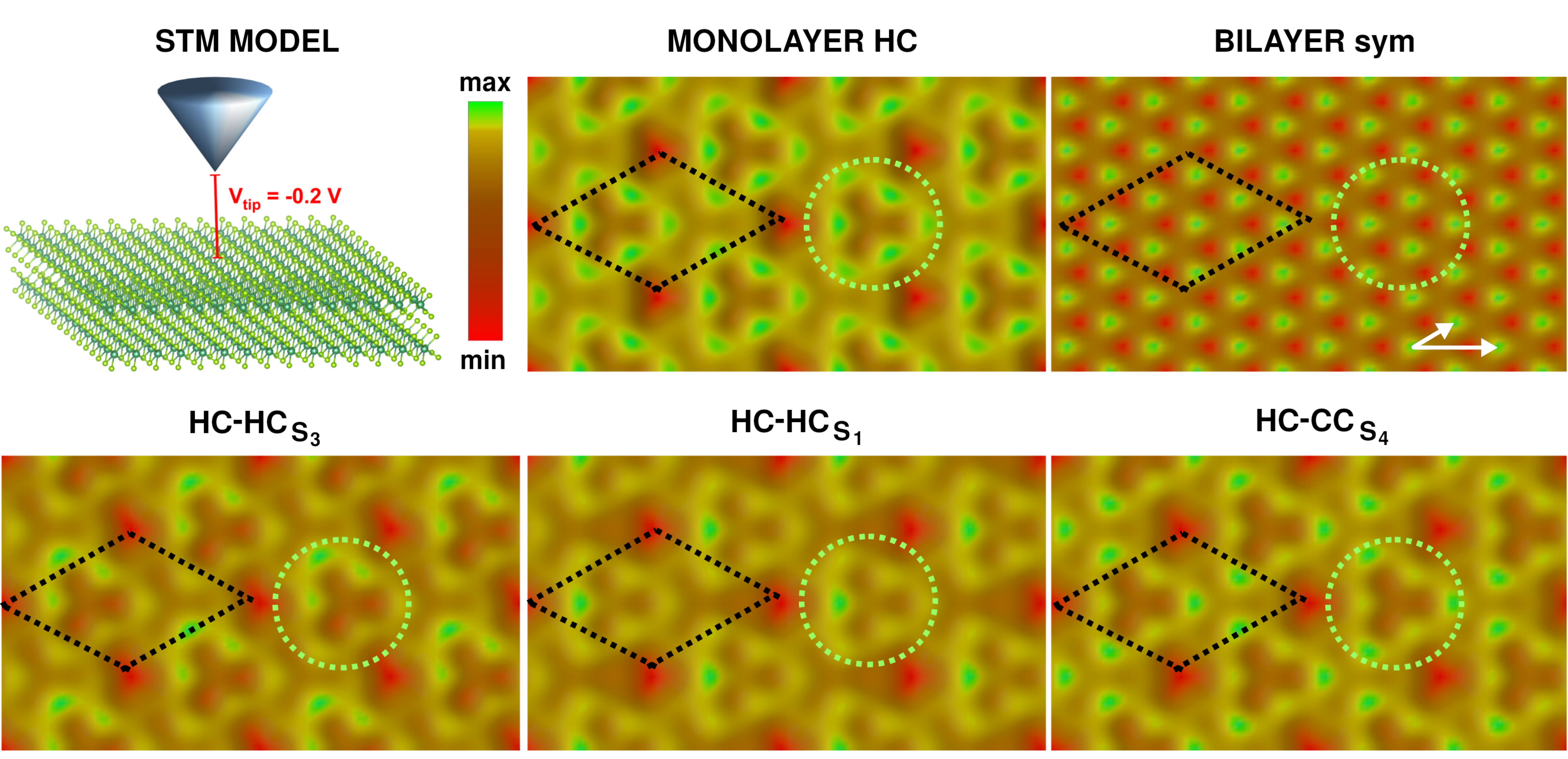}
	\caption{Simulated constant-current STM images from GGA+DF calculations of the 2H-NbSe$_2$ bilayer in the symmetric phase, as well as in the three most favourable CDW structures, namely HC--HC$_{S_3}$, HC--HC$_{S_1}$ and HC--CC$_{S_4}$. Data for the 1H-NbSe$_2$ monolayer in the HC structure are also shown, alongside a schematic view of the modeled STM setup. Black dashed rhombus and green dashed circle respectively denote the surface unit cell and the three-lobed patch. Note that the top layer of the three bilayers is always assumed to be a HC-type CDW. The white arrows are used for the discussion of the geometric structure factors.}
\label{fig:STM}
\end{figure*}

To check if it is possible to resolve the subtle interplay of the various patterns present in the CDW state, we employ our electronic structure calculations to model STM experiments. 
Theoretical constant-current STM images for the normal phase and the three most favorable CDWs identified in the 2H-NbSe$_2$ bilayer are shown in Figure~\ref{fig:STM}, as calculated in GGA+DF.
For reference, we also include data for the HC structure of the 1H-NbSe$_2$ monolayer, where we identify the three-lobed patch in the dashed circle.
High-symmetry of large apparent height (green) and small apparent height (red) patches defining the period of the CDW are clearly noted in the simulated constant-current STM images.
As visible in the bottom panels, these features become rather different for the three possible CDWs simulated for the 2H-NbSe$_2$ bilayer. 
First, both images for the HC--HC blend exhibit an asymmetric pattern. Second, the image for HC--CC$_{S_4}$ shows that the green spots at the tips of the three-lobed patch have a different size with respect to the monolayer data.
Overall, green spots are found at different positions for the three considered CDWs: on one of the protuberances of the three-lobed patch for HC--HC$_{S_3}$, on one of its dimples for HC--HC$_{S_1}$ and on all tips for HC--CC$_{S_4}$.
To understand the role played by a variation of the inter-layer distance, it is useful to compare the data obtained by GGA+DF with those obtained by pure GGA, which are reported in Appendix~\ref{sec:morestm}.
We observe that the CDW structures still exhibit patterns with a different symmetry, although the contrast between the various features seems less marked, owing to the increased separation between the two composing monolayers. This offers strong support to the analysis provided in this paragraph, which is unlikely to be affected by inaccuracies associated to the underlying DFT calculations.
 
 
To quantify the observed STM contrasts in Figure~\ref{fig:STM}, we calculate the corrugation values of the constant-current contours, i.e. the apparent height difference between green (max) and red (min) parts of the images. We find that the symmetric bilayer with no CDW exhibits the smallest corrugation (13 pm), the bilayers in the CDW state have an intermediate corrugation (29 to 39 pm), whereas the monolayer HC structure shows the largest corrugation (51 pm). We also calculate the apparent height differences of the green parts of the images themselves, which show the most detectable features of the different CDW blends in the NbSe$_2$ bilayer in comparison to the monolayer HC structure. We find that the apparent height differences of the green areas follow the same trend as the corrugations of the constant-current contours: the symmetric bilayer with no CDW exhibits the smallest value (2 pm), the bilayers in a CDW state have intermediate values (4 to 5 pm), and the monolayer HC structure shows the largest value (6 pm).
The identified trends from these apparent height difference values at the given tunneling condition are also transferable to other sets of tunneling parameters, since the corrugation and the contrast (apparent height differences) of constant-current contours follow an exponential decay with the tip-sample distance~\cite{chen2021introduction,PhysRevB.87.024417}. Increased current values are expected to result in the experimental differentiation between the CDW stacking configurations using STM equipments with a vertical resolution of better than 10 pm at low bias voltages (see also Appendix~\ref{sec:morestm}). This resolution has already been achieved in technology and has also been applied to the study of CDWs~\cite{renner1}, including bulk 2H-NbSe$_2$
\cite{[{}][{; details disclosed via private communication.}]renner2}.
Therefore, our analysis provides a theoretical guide for the experimental STM detection of distinguished CDW blend structures in 2H-NbSe$_2$, which can be also employed for similar studies of the stacking of other 2D materials.

Although the STM images shown in the bottom panels of Figure~\ref{fig:STM} are all obtained for the same CDW structure in the top monolayer (HC), our results show a certain sensitivity to the modulations in the bottom monolayer as well.
Therefore, we suggest that in clean samples, STM should be able to detect the changes associated to the inter-layer stacking. 
Naturally, defects are present in samples, thus the differences in relative intensity between the STM images may be more subtle and elusive to instruments that we would hope for. Although this is something that the rapid improvement in techniques for growth and measurement may overcome in a few years, it is instructive to provide a complementary analysis based on a different approach. To this aim, in the next subsection, we simulate geometric structure factors which can be used to interpret experimental data from X-ray diffraction, transmission electron microscopy or electron energy loss spectroscopy.

\subsection{Charge density and structure factors}
 The charge distributions of the modelled systems are computed integrating the electronic states over the occupied Nb band, down to \SI{-0.4}{\electronvolt} below the Fermi level. 
 From the three-dimensional object thus produced, we obtain geometric structure factors $f(h,k,l)$ for the 2H-NbSe$_2$ bilayer, which have the advantage of factoring out the in-plane displacement, thus highlighting the symmetry of the lattice and its charge distribution. By treating separately lattice and charge modulations, we can assess their roles in the transition from one symmetry to another~\cite{VESTA-JACr2011}. In general, this analysis can also provide a tool to detect changes of symmetry not occurring at the surface but in lower layers.

\begin{figure}
\centering
\includegraphics[trim= 0cm 0cm 0cm 0cm, width=1\linewidth]{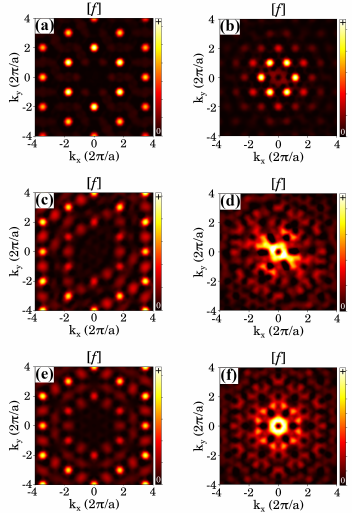}
	\caption{Moduli of the geometric structure factors of the 2H-NbSe$_2$ bilayer in the plane $(h,k,0)$ calculated from the structural data only (left panels; the wavelength is 1.2 \AA) or directly from the electronic charge arising from states in the vicinity of the Fermi level (right panels; see main text for more details). (a)-(b) intensity map for the HC--HC$_{S_3}$ structure; (c)-(d) intensity map for the HC--HC$_{S_3}$ structure minus the symmetric solution without CDW ; (e)-(f) intensity map for the HC--CC$_{S_4}$ structure minus the symmetric solution without CDW.}
\label{fig:str-FTs}
\end{figure}


 We select vectors in the $(h,k,l)$ plane with $l=0$ to retain the in-plane electron momentum. In principle, any constant $l$ would do that, but for $l \neq 0$ an additional high peak appears at $(0,0,l)$.
 We first calculate geometric structure factors for X-ray diffraction from structural data only, in the approximation of non-interacting atoms~\cite{VESTA-JACr2011,Aguilar-Marin_2020}. These plots are reported on the left hand side of Figure~\ref{fig:str-FTs}: panel (a) shows data for the ground state system, panel (c) shows its difference with respect to the fully-symmetric (no CDW) structure, and panel (e) shows the difference between the excited state HC--CC$_{S_4}$ and the fully-symmetric structure.
 As expected, the moduli of the structure factors are characterised by high Bragg peaks; in each plane with constant $l$, these peaks are found at $(\pm \sqrt{3}, \pm 1)$, $(0, \pm 2)$, in units of $2\pi/a$, where $a$ is the unit in-plane lattice constant of the primitive unit cell.
Due to the large intensity of the Bragg peaks from the structural model, the CDW peaks appear very dim in the modulus of the structure factors, even when we filter the symmetric structure out, as in Figure~\ref{fig:str-FTs}(c). Nevertheless, the difference patterns shown in panels (c) and (e) have a very different shape, which suggests that experimental techniques suitable for determining the periodic lattice distortions directly should be able to distinguish among different stacking sequences, if sufficiently accurate.

More detailed information can be obtained by inspecting the geometric structure factors calculated directly from the charge density.
These terms are proportional to the Fourier transforms of the charge density (or, more in general, scattering density) integrated over the unit cell~\cite{VESTA-JACr2011,Aguilar-Marin_2020,warren1969warren}. By considering only the charge density arising from the states in the vicinity of the Fermi energy, we obtain a term of comparison for the Fourier transforms of experimental data from STM. The calculated plots are reported on the right hand side of Figure~\ref{fig:str-FTs}, where panel (b) shows the data for the ground state system, panel (d) shows its difference with respect to the fully-symmetric (no CDW) structure, and panel (f) shows the difference between the excited state HC--CC$_{S_4}$ and the fully-symmetric structure.
The scenario outlined by the structure factors from the partial charge density is slightly different from the one arising from the structural data only. Figure~\ref{fig:str-FTs}(b) shows that the signal is dominated by peaks at $(\pm 1/\sqrt{3}, \pm 1)$ and $(\pm 2/\sqrt{3}, 0)$. These reciprocal lattice vectors correspond to the direct-space vectors of $(3/2, \sqrt{3}/2)$, which result from the second nearest neighbours in the lattice, as indicated by the white vectors in Figure~\ref{fig:STM}. The CDW peaks are now more visible and are located at 1/3 of the distance between the central point and the first order Bragg peaks. The latter are dimmer than those observed in Figure~\ref{fig:str-FTs}(a), but are still visible.
When inspecting the difference patterns in panels (d) and (f), we notice that the dominant contribution is due to the CDW peaks, which are found in the tips of a six-fold star. This is particularly evident for the HC-CC$_{S_4}$ solution, which features a higher symmetry with respect to the ground state.
The CDW peaks are also accompanied by their corresponding second neighbor contribution, distributed along the innermost hexagon allowed by the supercell size (see also Figure~\ref{fig:raw_FTs}).
The patterns shown in Figure~\ref{fig:str-FTs}(d) and Figure~\ref{fig:str-FTs}(f) show again marked differences, which suggests that they can be useful for experimental analysis if a sufficient accuracy is achieved.

Overall, the structure factors calculated from the partial charge density differ markedly from those from the sole structural data, which reflect X-ray diffraction experiments. Such difference, in the peak intensities and positions, originates from the nature of the charge in the integrated range, whose $l$-character is known to be 2. The symmetry of the geometric structure factors reflects the peculiar symmetry present in the CDW blends and provides a valuable term of comparison for experiments based on X-ray diffraction~\cite{burk1992}, STM~\cite{ayanipreprint2024,sivakumar2024influence}, and TEM~\cite{wang23tem}, to be used in alternative to, or in combination with, the direct STM simulations presented in the previous section.

\section{Discussion and Conclusions}
The results outlined in the previous sections demonstrate that the inter-layer stacking may play an important role on the physics of the 2H-NbSe$_2$ bilayer, changing the symmetry of the ground-state with respect to the 1H-NbSe$_2$ monolayer. Due to the higher number of degrees of freedom, a plethora of complex excited states form, each of them characterized by subtle differences in the electronic excitations. The most visible ones are peaks derived from hybridized $p_z$ states of Se atoms belonging to the layers connecting the two monolayers. 
Around the Fermi level, the most favourable CDW structures exhibit a similar density of carriers, but more evident effects are expected to appear in the transport properties, especially in light of the expected anisotropy. The latter becomes particularly evident in the results obtained from the STM simulation, where the combination of the different modulations leave interesting fingerprint that do seem in reach for experimental detection. Similar conclusions can be reached from the analysis of the geometric structure factors, which can be connected to direct experiments of X-ray diffraction or to the Fourier transforms of data from STM or TEM. Further insight into this problem can also be obtained from high-resolution data from non-contact atomic-force microscopy (AFM)~\cite{ayanipreprint2024,sivakumar2024influence} and scanning transmission electron microscopy (STEM)~\cite{martis2023imaging,cheng24nanolet}. Therefore, we encourage experimental groups to consider looking for fingerprints of the predicted stacking, which is likely to appear in mixed grain boundaries, as reported in previous literature. Since many studies were focused on the 2H-NbSe$_2$ surfaces and discussed in terms of theoretical data for the 1H-NbSe$_2$ monolayer, we should also think of re-evaluating former interpretations by assuming a non-negligible role of the stacking on the measured quantities.

Although our work has been focused on the 2H-NbSe$_2$ bilayer, we do not expect that the qualitative conclusions summarized above will change for the bulk, which is composed of stacked bilayers. While an increased number of degrees of freedom may allow for the construction of more complex blends and displacements, their effect on the physical properties of the system will not be as marked as the one associated to the bilayer formation. After many investigations on TMDs of the 1T-type, our results demonstrate that the inter-layer stacking plays a role in the 2H-type as well. Albeit no dramatic effects such as metal-to-insulator transitions are observed, we do expect fingerprints of the stacking to emerge with the constantly improving opportunities offered by modern facilities for microscopy and spectroscopy.
 
Concerning possible applications or interesting points of analysis for future work, the lower symmetry observed in the most stable CDW structures can be interesting for magnetism, as shown in many reports so far~\cite{flicker-PRB.92.201103,myself-NPGAM2020,myself-PRB.98.195419,wickramaratne-PRX.10.041003,wickramaratne-PRB.104.L060501}.
The sensitivity of bilayers suggests that vibrational modes could be used in combination with other external stimuli to tune collective electronic states. In this regard, it has been pointed out that external pressure can tune the types and functionalities of CDWs. Furthermore, we also note that instead of in-plane displacements, addressed in this work, bilayers can feature other geometrical configurations, such as layers twisting with suitable angles~\cite{Goodwin_2022,PhysRevB.108.224111,lastreff}.
 Naturally, the interaction between layers will play a similar role in the combination of lattice distortions and charge modulations to promote or hinder collective excitations such as superconductivity.\\
 
Regarding electronic structure calculations, future theoretical work may be focused on understanding the connection between Raman modes and inter-layer stacking, in relation to both blend and displacement. Previous studies have suggested that the vertical alignment of CDWs does not cause a discontinuity in the frequency of the shear mode, see e.g. Figure~4(b) in ref.~\cite{he_r-2DMat2016}. Although the authors interpreted this feature as a hint of a vertical alignment of the CDWs, likely induced by defects, this does not seem possible for a 2H type, due to the particular layer stacking illustrated in Figure~\ref{fig:blend-displcm_summary}(a). 
 Interpreting Raman data will require understanding the variations of phonon spectra across the various CDW structures. Although this task involves several methodological difficulties, it could help identify what are the key signatures of of CDW stacking, possibly even separating blend and displacements. In addition, such calculations would make it possible to evaluate the corrections associated to the vibrational entropy and to understand their role in the energy hierarchy of the inter-layer stacking.
  
%

\section*{Acknowledgments}
The authors are thankful to S.\ Leb\`egue, M.\ I.\ Katsnelson, F. Flicker, I.\ I.\ Mazin, D.\ D.\ Sarma, D. Kepaptsoglou, L.\ C.\ Rhodes, C.\ A.\ Marques, V.\ K.\ Lazarov, I.\ E.\ Brumboiu, J.\ A.\ Do Nascimento, O. Gr{\aa}n\"as, P.\ D.\ C.\ King and \'Arp\'ad P\'asztor for fruitful discussions. The computational work was enabled by resources provided by the Swedish National Infrastructure for Computing (SNIC) at the Center for High Performance
Computing (PDC) in Stockholm, Sweden, partially funded by the Swedish Research Council through Grant Agreement No. 2018-05973. Further computational resources were provided by
the Korean Institute of Science and Technology Information (KISTI). This study was supported by the National Research Foundation (NRF) funded by the Ministry of Science of Korea
-- Grants No. 2017R1D1A1B03033465 and 2022R1I1A1A01071974. I.\ D.M.\ also acknowledges support from the European Research Council (ERC), Synergy Grant FASTCORR, Project No. 854843,
and from the JRG Program at APCTP through the Science and Technology Promotion Fund and Lottery Fund of the Korean Government, as well as from the Korean Local Governments,
Gyeongsangbuk-do Province and Pohang City. K.P. acknowledges the NRDIO-Hungary grant no. FK124100 and a Bolyai Fellowship of the Hungarian Academy of Sciences. A.\ A.\ also
acknowledges financial support from the German Research Foundation within the bilateral NSFC-DFG Project No.\ ER 463/14-1.
This research is part of the project No. 2022/45/P/ST3/04247 co-funded by the National Science Centre and the European Union's Horizon 2020 research and innovation programme under the Marie Sk{\l}odowska-Curie grant agreement no. 945339.

\section*{CRediT authorship contribution statement}
{\bf{F. Cossu:}} Formal analysis, Funding acquisition, Investigation, Validation, Visualization, Writing - original draft, Writing - review \& editing.
{\bf{D. Nafday:}} Data curation, Investigation, Methodology, Validation, Writing - original draft, Writing - review \& editing.
{\bf{K. Palot\'as:}} Formal analysis, Software, Visualization, Writing - original draft, Writing - review \& editing.
{\bf{M. Biderang:}} Software, Writing - review \& editing.
{\bf{H.-S. Kim:}} Methodology, Resources, Software, Writing - original draft, Writing - review \& editing.
{\bf{A. Akbari:}} Conceptualization, Funding acquisition, Supervision, Writing - original draft, Writing - review \& editing.
{\bf{I. Di Marco:}} Conceptualization, Funding acquisition, Methodology, Project administration, Resources, Software, Supervision, Visualization, Writing - original draft, Writing - review \& editing.

\appendix
\section{Additional STM simulations}\label{sec:morestm}
\begin{figure*}[t]
\centering
\includegraphics[trim= 0cm 0.0cm 0cm 0cm, width=1.0\linewidth]{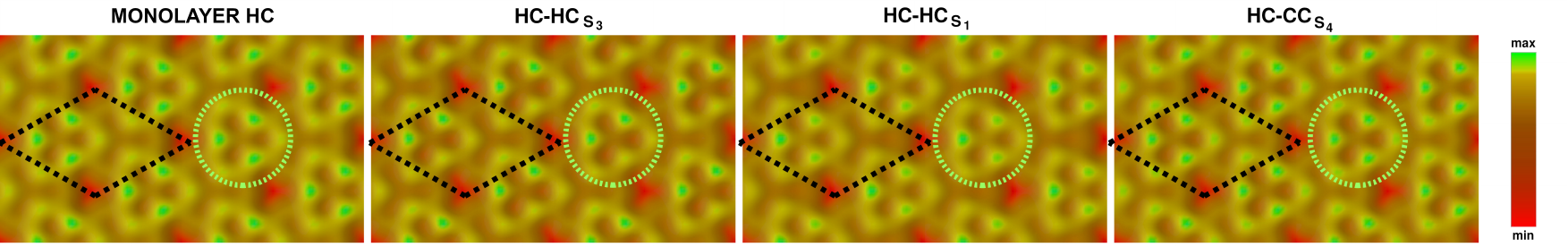}
	\caption{Simulated STM images using GGA only, without vdW corrections. The leftmost panel shows the map for the monolayer 1H-NbSe$_2$ in the HC state. The other three panels show the maps for the most favourable states identified for the 2H-NbSe$_2$ bilayer, namely HC--HC$_{S_3}$, HC--HC$_{S_1}$ and HC--CC$_{S_4}$. As shown in the color map on the right, green and red spots represent high and low apparent heights of constant current contours, respectively.}
\label{fig:STMgga}
\end{figure*}
Figure~\ref{fig:STMgga} shows STM images obtained from GGA calculations without vdW corrections. Note that the contrast of the various features is slightly less marked with respect to the data obtained with GGA+DF, reported in Figure~\ref{fig:STM}. These differences are noticeable also for the monolayer, so they also reflect a change induced between the Nb and Se layers. Overall, the fundamental differences in the symmetry of the images for the different CDWs are still visible.
With vdW correction (Figure~\ref{fig:STM}) the maxima are distributed almost equally on the protuberances and dimples of the three-lobed patch, whereas without vdW corrections (Figure~\ref{fig:STMgga}) the maxima are larger at the dimples. A similar change is observed for the images of the bilayer, especially for HC--HC$_{S_3}$ and HC--HC$_{S_4}$, where maxima shift from the dimples to the protuberances when vdW corrections are added.

\begin{figure}[b]
\centering
\includegraphics[trim= 0cm 0.0cm 0cm 0cm, width=\linewidth]{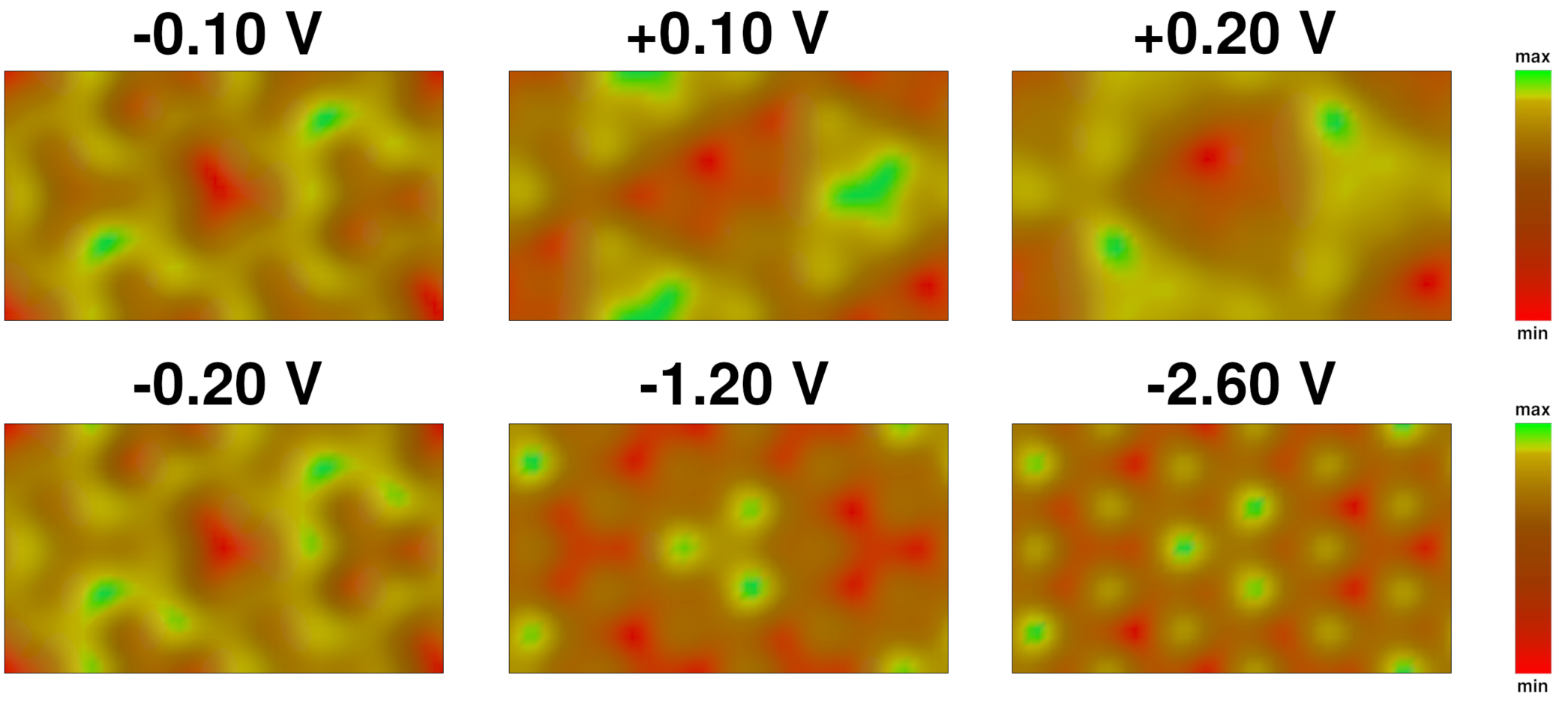}
	\caption{Simulated constant-current STM images of the HC--HC$_{S_3}$ CDW structure of the 2H-NbSe$_2$ bilayer, using the same computational settings used for Figure~\ref{fig:STM}, but varying the bias voltage.}
\label{fig:stm_V_dep}
\end{figure}

To illustrate the dependence of the theoretical STM images on the bias voltage, we performed a series of calculations for the ground state HC--HC$_{S_3}$. The resulting constant-current STM images are shown in Figure~\ref{fig:stm_V_dep}. The asymmetry in the STM contrast, amply discussed in the manuscript, is visible at low voltages, while a change of contrast occurs only at large negative voltages. These data demonstrate that the results discussed in the paper do not depend on a very restricted set of computational parameters.

\section{Raw data for the structure factors}\label{sec:rawdata}
\begin{figure}[t]
\centering
\includegraphics[trim= 0cm 0.0cm 0cm 0cm, width=0.9\linewidth]{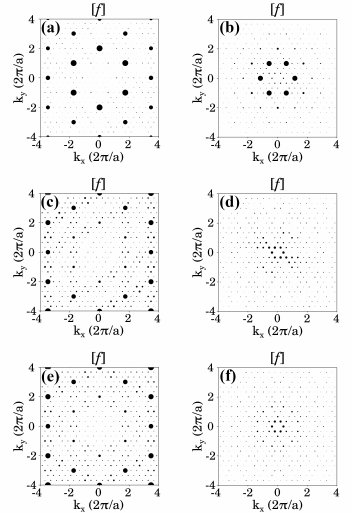}
	\caption{Moduli of the geometric structure factors of the 2H-NbSe$_2$ bilayer in the plane $(h,k,0)$ calculated from the structural data only (left panels; the wavelength is 1.2 \AA) or directly from the electronic charge arising from states in the vicinity of the Fermi level (right panels; see main text for more details). (a)-(b) intensity map for the HC--HC$_{S_3}$ structure; (c)-(d) intensity map for the HC--HC$_{S_3}$ structure minus the symmetric solution without CDW ; (e)-(f) intensity map for the HC--CC$_{S_4}$ structure minus the symmetric solution without CDW. The size of the circles is proportional to the intensity of the signal, in arbitrary units.}
\label{fig:raw_FTs}
\end{figure}

For illustrative purposes, the geometric structure factors presented in Figure~\ref{fig:str-FTs} consisted of color maps, obtained through a cubic interpolation from the raw data calculated by VESTA. For completeness, the raw data are reported in Figure~\ref{fig:raw_FTs}. The discrete mesh defining the sampling of the reciprocal space is dictated by the size of the supercell used for the calculations.

\end{document}